\documentclass[aps,prd,twocolumn,superscriptaddress]{revtex4-2}
\usepackage{graphicx}
\usepackage{amsmath}
\usepackage{amssymb}
\usepackage{hyperref}
\usepackage{xcolor}
\usepackage{array}
\usepackage{booktabs}
\usepackage{subfigure} 
\usepackage{subcaption}
\usepackage{float}
\usepackage{tikz}
\usepackage{xcolor}
\usepackage{comment}

\begin{document}

\title{Cosmic Trajectories calculation with state of the art lattice QCD equation of state}

\author{Lorenzo Formaggio}\thanks{Corresponding author: lformagg@cougarnet.uh.edu}
\affiliation{Department of Physics, University of Houston, Houston, TX 77204, USA }

\author{Francesco Di Clemente}
\affiliation{Department of Physics, University of Houston, Houston, TX 77204, USA }

\author{Geetika Yadav}
\affiliation{Department of Physics, University of Houston, Houston, TX 77204, USA }

\author{Alessandro Drago}
\affiliation{INFN, Sezione di Ferrara, via Saragat 1, I-44122 Ferrara, Italy}
\affiliation{Department of Physics and Earth Science, University of Ferrara, via Saragat 1, I-44122 Ferrara, Italy}

\author{Claudia Ratti}
\affiliation{Department of Physics, University of Houston, Houston, TX 77204, USA }
\date{\today}
\begin{abstract}
We compute the full cosmic trajectories of the early Universe across the QCD phase diagram as the plasma cools from $T\simeq500\,$MeV to $30\,$MeV, assuming $\beta$-equilibrated matter.  
The trajectories are obtained by simultaneously solving baryon-number, electric-charge, and lepton-asymmetry conservation, closed by a state-of-the-art lattice-QCD equation of state: a fourth-order Taylor expansion in the chemical potentials that merges the latest $(2\!+\!1)$-flavor susceptibilities with charm-quark contributions, thus delivering a consistent $(2\!+\!1\!+\!1)$-flavor equation of state.  
Results are compared with an ideal quark-gluon plasma and with a hadron-resonance gas to highlight interaction effects. 
Two cases of primordial lepton asymmetries are analyzed: a symmetric configuration $(\ell_e=\ell_\mu=\ell_\tau=\ell/3)$ and an asymmetric one $(\ell_e=0,\;\ell_\mu=-\ell_\tau)$. 
Increasing $|\ell|$ systematically drives the trajectories toward larger values of $\mu_B$ and more negative $\mu_Q$. In the asymmetric case, a non-monotonic ``bounce’’ develops when the $\tau$ chemical potential reaches $m_\tau$, generating a maximum in $\mu_B(T)$, the position of which depends on $\ell_\tau$. Assuming a modest $\mu_{Q}$-dependence of the lattice-QCD critical end point estimates (obtained at $\mu_{Q} = 0$), the trajectories for all lepton asymmetries explored ($|\ell|\lesssim 0.1$) lie to their left, implying that in a standard cosmological scenario the QCD transition is almost certainly a smooth crossover. Nevertheless, we estimate the magnitude of baryon and lepton asymmetries needed to obtain a cosmic trajectory closer to the QCD critical point, providing inputs for future studies of the strong-interaction epoch.
\end{abstract}
\maketitle

\section{Introduction}

Within the framework of the hot Big Bang cosmological model, the thermal history of the early Universe includes at least two major epochs during which phase transitions associated with fundamental interactions are believed to have occurred. The first one is the electroweak (EW) transition, occurring at temperatures of the order of $T_{\mathrm
{EW}}\sim100$ GeV, during which the Higgs mechanism spontaneously broke the $SU(2)_L\times U(1)_Y$ gauge symmetry of the Standard Model (SM), endowing elementary fermions and EW gauge bosons (W, Z) with mass. The second is the Quantum Chromodynamics (QCD) phase transition, which took place at significantly lower temperatures, around $T_\mathrm{QCD}\sim150$ MeV, as the Universe cooled to a point where deconfined quarks and gluons underwent confinement into color-neutral hadronic bound states.

Within the Standard Model of particle physics, both the EW and QCD transitions are understood to be continuous crossovers rather than genuine phase transitions, based on non-perturbative studies such as lattice gauge theory. 
Nevertheless, a growing body of observational and theoretical evidence from cosmology and high-energy physics strongly motivates the existence of physics beyond the Standard Model. In particular, numerous scenarios — including minimal extensions such as the addition of a scalar singlet — can render the EW transition first-order \cite{Espinosa:1993bs}, which has significant implications for mechanisms like EW baryogenesis.

In contrast, constructing viable extensions of the SM that induce a first-order QCD phase transition is more challenging, and far fewer mechanisms have been proposed in the literature to realize such a scenario \cite{Iso:2017uuu,Hambye:2018qjv}. Despite this, a strongly first-order QCD transition is of substantial interest, as it could generate observable cosmological signatures. Specifically, a first-order QCD transition would proceed via nucleation of bubbles of the confined phase in a background of deconfined quark-gluon plasma, leading to the production of a stochastic background of gravitational waves through bubble collisions, turbulence, and sound waves in the early Universe plasma \cite{Schwarz:2003du,PhysRevD.30.272}. Moreover, the detailed dynamics and nature of the QCD transition have profound consequences for the formation and abundance of exotic relics from the early Universe. One particularly intriguing consequence is the possibility of enhanced primordial black hole (PBH) formation during a first-order QCD transition. As the equation of state temporarily softens and the speed of sound decreases near the transition, density perturbations may undergo gravitational collapse more efficiently, leading to the creation of PBHs with mass scales set by the horizon size at $T_{QCD}$ \cite{Jedamzik:1996mr,Jedamzik:1999am,Byrnes:2018clq,Vovchenko:2020crk,Bodeker:2020stj}. Such black holes are compelling candidates for dark matter, and their potential contribution to the dark matter abundance remains a subject of active theoretical and observational investigation. Another possibility is that the bubbles generated by the first-order phase transition lead to stable strange quark matter droplets (strangelets), which also remain viable dark matter candidates \cite{Sinha:2022jfr,DiClemente:2024lzi}.
The possibility that the Universe went through a first-order phase transition during the QCD epoch was recently explored in Ref. \cite{Gao:2021nwz} using an equation of state obtained with functional methods, and in Ref. \cite{Ferreira:2025zeu} in the context of the quark-meson model. In both cases, large lepton asymmetries were considered. In fact, while the baryon-antibaryon asymmetry is very well constrained in our Universe, a larger uncertainty in the lepton ones leads to a larger region spanned in the QCD phase diagram.

Let us review the state-of-the-art constraints on these asymmetries. Despite the expectation that the Big Bang should have produced matter and antimatter in equal quantities, our Universe exhibits a striking baryon–antibaryon asymmetry, with no significant relic abundance of antimatter detected in cosmic structures. This asymmetry is conventionally quantified in terms of conserved quantum numbers, specifically the baryon number $B$, lepton number $L$, and electric charge $Q$, which are expected to be conserved in most low-energy physical processes. The baryon asymmetry of the Universe is rigorously defined by the ratio of the net baryon number density $n_B$ to the entropy density of the Universe $s$, $b=n_B/s$. Observational constraints, primarily from measurements of the cosmic microwave background (CMB) anisotropies and Big Bang Nucleosynthesis (BBN), determine this quantity to be $b=(8.60\pm0.06)\times10^{-11}$ \cite{gao2022cosmology,wygas2018cosmic}.
Within the framework of the Standard Model (SM) of particle physics, the generation of such a small but non-zero baryon asymmetry of the Universe is not achievable. While the SM includes some of the necessary Sakharov conditions—namely baryon number violation via nonperturbative EW sphaleron processes, C and CP violation through the CKM matrix, and a departure from thermal equilibrium during the EW phase transition—these ingredients are insufficient in magnitude to reproduce the observed value of $b$ \cite{Sakharov:1967dj}. In particular, the strength of CP violation and the nature of the EW phase transition in the SM are inadequate, necessitating physics beyond the Standard Model to generate the observed asymmetry \cite{Barr:1979ye,Farrar_1994}.

A compelling mechanism that addresses this issue is leptogenesis, which posits that a primordial asymmetry in the lepton sector is first generated, typically through the out-of-equilibrium decays of heavy right-handed Majorana neutrinos in the context of the seesaw mechanism. The resulting lepton number violation and CP-violating interactions produce a non-zero lepton asymmetry $\ell=n_L/s$, which is partially converted into a baryon asymmetry through EW sphaleron processes that violate $B+L$ while conserving $B-L$. Under the assumption of thermal equilibrium and efficient sphaleron dynamics, the SM predicts a linear relation between the baryon and lepton asymmetries, given by $\ell=-\frac{51}{28}b$ \cite{PhysRevD.42.3344}, reflecting the redistribution of asymmetries among fermionic degrees of freedom.

Nevertheless, there exist several theoretical approaches beyond the SM that allow for a much larger present-day lepton asymmetry, characterized by $|\ell|\gg b$, with no theoretical bias toward a particular sign of $\ell$. Such large lepton asymmetries can arise in scenarios where the sphaleron processes are either dynamically suppressed or rendered ineffective by occurring after the EW symmetry breaking phase transition. For instance, in models involving low-scale leptogenesis, Affleck-Dine baryogenesis, or sterile neutrino oscillations in the early Universe, the lepton asymmetry may be generated at temperatures below the sphaleron freeze-out threshold, thereby avoiding significant conversion to baryon number \cite{Eijima:2017anv}. These possibilities have important implications not only for baryogenesis but also for neutrino physics, dark matter production, and the thermal history of the early Universe.

From an observational standpoint, the magnitude of the lepton asymmetry remains only weakly constrained relative to the baryon asymmetry. Although the requirement of overall electric charge neutrality in the Universe imposes a strong correlation between the asymmetries of charged leptons and baryons — thereby limiting the charged lepton asymmetry to be of the same order of magnitude as the baryon asymmetry $b\sim10^{-10}$ (see \cite{Caprini:2003gz} for a quantitative upper bound) — this constraint does not apply directly to electrically neutral particles. As a result, a significant lepton asymmetry could be sequestered in the neutrino sector without violating observational constraints on charge neutrality or the baryon-to-photon ratio.

In particular, the presence of a large neutrino-antineutrino asymmetry is theoretically viable and may have cosmological implications, especially during epochs such as BBN and recombination. The lepton asymmetry stored in relic neutrinos contributes to the energy density of the early Universe and alters the effective number of relativistic degrees of freedom, $N_\mathrm{eff}$, thereby affecting both the abundance of light nuclei synthesized during BBN and the anisotropy spectrum of the CMB. Current bounds derived from Planck satellite observations of the CMB anisotropies place a model-independent constraint on the lepton asymmetry at the level of $|\ell|<0.012$ at 95\% confidence level \cite{Oldengott:2017tzj}. These results are consistent with independent analyses based on BBN, which examine the influence of neutrino degeneracy parameters on primordial element abundances and yield compatible limits on the neutrino asymmetry \cite{Mangano:2011ip}.
These relatively loose bounds leave open the possibility that the lepton asymmetry, particularly in the neutrino sector, could be several orders of magnitude larger than the baryon asymmetry.

At high baryon densities and low temperatures, first-principle lattice techniques are not presently feasible, due to the fermion sign problem. However, several models predict a critical point on the QCD phase diagram, separating the crossover transition at small densities from a first-order phase transition occurring as the density/chemical potential increases (for a recent review see \cite{Bzdak:2019pkr,Du:2024wjm}). While in the past the theoretical predictions for the location of the QCD critical point were scattered all over the QCD phase diagram, more recent ones seem to concentrate around a considerably narrower region, with chemical potentials in the range 450 MeV $\leq\mu_B\leq650$ MeV \cite{Fu:2019hdw,Gunkel:2021oya,Gao:2020fbl,Hippert:2023bel,Basar:2023nkp,Clarke:2024ugt,Shah:2024img}. In most of these cases, the location is on the pure $(T,~\mu_B)$ plane, with $\mu_S=\mu_Q=0$.

In this manuscript, we examine the influence of an undetermined lepton asymmetry on the evolution of the Universe through the QCD phase diagram.  In Section II, we formulate the three relevant conservation conditions for the early Universe: baryon-number conservation, electric-charge neutrality, and individual lepton-flavor asymmetries. We specify the three equations of state employed: an ideal quark–gluon plasma, the state-of-the-art fourth-order Taylor-expanded lattice QCD equation of state with 2+1+1 flavors, and the hadron-resonance gas model. Section III presents our numerical results, explicitly calculating cosmic trajectories for symmetric and asymmetric lepton-asymmetry scenarios, given different lepton asymmetry values. We compare these trajectories with the ideal-gas of quarks and gluons and the hadron-resonance gas baselines. Finally, in Section IV, we summarize our main findings and discuss their implications concerning how closely the early Universe's evolution may approach the hypothesized QCD critical endpoint.

\section{Theory}
In this section, we outline the main features of the method employed to compute the cosmic trajectories of the early Universe, following the approach proposed in Ref. \cite{wygas2018cosmic}. The procedure involves defining a system of equations that depends on temperature and the relevant chemical potentials. An equation of state, either that of an ideal quantum gas or one derived from lattice QCD, is needed to close the system of equations. 
By solving this system at fixed temperature values, we can reconstruct the cosmic trajectory point by point in the QCD phase diagram.
\subsection{Early Universe Conditions}
\label{Early Universe Condition}

Based on the discussion in the introduction, we can assume the conservation of baryon asymmetry, of lepton asymmetries, and the charge neutrality of the Universe. The temperature regime we are investigating is 30 MeV $\lesssim T \lesssim$ 500 MeV, as we know that the pseudo-critical temperature at small chemical potentials is around $T_0\sim160$ MeV \cite{Borsanyi:2020fev}. At $T_\mathrm{baryon}\sim 10$ TeV, baryogenesis occurs, fixing the baryon asymmetry $b=8.6\cdot 10^ {-11}$ \cite{gao2022cosmology,wygas2018cosmic} and inferred by \cite{refId0}. This leads to Eq. (\ref{bcons}), where $n_i^\mathrm{asym}(T,\mu)=n_i(T,\mu)-n_i(T,-\mu)$ is the net number density of particles of species $i$, and $B_i$ is the baryon fraction brought by each particle (for quarks, $B_{quark}=\frac{1}{3}$). $s$ is the total entropy density of the Universe. The Universe is electrically neutral; therefore, we introduce Eq. (\ref{qneutr}), which expresses the charge neutrality condition, where $Q_i$ denotes the fractional electric charge of each particle species. The last set of equations comes from the fact that, as already discussed, as long as we are at temperatures larger than $T_\mathrm{osc}\simeq10$ MeV, neutrino oscillations are not efficient. Then we can assume the conservation of the three lepton asymmetries \cite{mangano2011constraining}, expressed in Eq. (\ref{lepasym}), where $i=e,\mu,\tau$. Our system of equations is therefore:

\begin{equation}
    b=\sum_i B_i \frac{n_i^{\mathrm {asym }} }{s}
    \label{bcons}
\end{equation}
\begin{equation}
     0=q =\sum_i Q_i \frac{n_i^{\mathrm {asym }}}{s} 
     \label{qneutr}
\end{equation}
\begin{equation}
\ell_i s=n_i^\mathrm{asym}+n_{\nu_i}^{\mathrm {asym }}. \label{lepasym}
\end{equation}

We are assuming five conserved quantities; this leads to introducing five independent chemical potentials:
\begin{equation}
    \{ \mu_B,\mu_Q, \mu_{\nu e}, \mu_{\nu \mu}, \mu_{\nu \tau}\}.
\end{equation}
Each particle contributes with its own chemical potential. Therefore, if we consider a Universe made of leptons and the four first generations of quarks, we have ten independent chemical potentials (six from the leptons and neutrinos, four from the quarks). To avoid this mathematical problem, we assume that everything is in beta equilibrium: the QCD phase transition is expected to last around $10~ \mu s$, while the weak interaction time scale is $\tau_\mathrm{weak}\sim10^{-2}~ \mu s$, which is fast enough to consider the beta processes at  equilibrium:
\begin{equation}
\begin{aligned}
f_i+\bar{f}_j & \rightleftharpoons \nu_{f_i}+\bar{\nu}_{f_j} \\
q_u+e^{-} & \rightleftharpoons q_d+\nu_e.
\label{beta equilibrium}
\end{aligned}
\end{equation}
This introduces the following relationships between the chemical potentials:
\begin{equation}
\begin{gathered}
\mu_\mu=\mu_e+\mu_{\nu_\mu}-\mu_{\nu_e} \\
\mu_\tau=\mu_e+\mu_{\nu_\tau}-\mu_{\nu_e} \\
\mu_{down}=\mu_{u p}+\mu_e-\mu_{\nu_e} \\
\mu_{down}=\mu_{strange }\\
\mu_{up}=\mu_{charm},
\end{gathered}
\end{equation}
thus effectively reducing the number of free variables from ten to five. We can write the baryon and electric charge densities explicitly:

    \begin{equation}
    n_B=\frac{1}{3}n_{up}+\frac{1}{3}n_{down}+\frac{1}{3}n_{strange}+\frac{1}{3}n_{charm}
    \label{barion density}
\end{equation}
    \begin{equation}
    n_Q=\frac{2}{3}n_{up}-\frac{1}{3}n_{down}-\frac{1}{3}n_{strange}+\frac{2}{3}n_{charm}
    \label{charge density}.
\end{equation}
The baryon chemical potential $\mu_B$, electric charge chemical potential $\mu_Q$, and strangeness chemical potential $\mu_S$ are then given by:
\begin{equation}
    \mu_B=\mu_{up}+2\mu_{down}
\end{equation}
\begin{equation}
    \mu_Q=\mu_{up}-\mu_{down}
\end{equation}
\begin{equation}
    \mu_S=0.
\end{equation}

Photons and gluons, since they are massless, have their chemical potential fixed to zero:
\begin{equation}
    \mu_\gamma=\mu_{g}=0; 
\end{equation}
between the lepton chemical potentials and the neutrino ones, the following relation holds:
\begin{equation}
    \mu_l=\mu_{l\nu}-\mu_Q.
    \label{lepton chem}
\end{equation}

So far we talked about including up, down and strange quarks, gluons, photons and all the leptons with their neutrinos. In principle, in the QGP phase we should consider all elementary particles of the standard model, reported in Table \ref{table:qgp}. However, we are investigating the Universe from $T\simeq 30$ MeV up to $T\simeq500$ MeV. In this regime, we can safely assume that all the particles heavier than the charm quark bring negligible contributions to the thermodynamics. For this reason, we are going to consider all leptons with their neutrinos, photons, the first four generations of quarks and gluons. We will actually investigate two cases: with and without the charm quark, in order to understand how its presence affects the nature of the transition and at what temperature its contribution becomes really sizeable.

\begin{table}
\begin{tabular}{|l|l|l|l|l|}
\hline 
Particle &  & mass & spin & $g$ \\
\hline 
Quarks & $t, \bar{t}$ & $173\,\mathrm{GeV}$ & $\frac{1}{2}$ & $2\cdot3$ \\
& $b, \bar{b}$ & $4\,\mathrm{GeV}$ & & \\
& $c, \bar{c}$ & $1\,\mathrm{GeV}$ & & \\
& $s, \bar{s}$ & $100\,\mathrm{MeV}$ & & \\
& $d, \bar{d}$ & $5\,\mathrm{MeV}$ & & \\
& $u, \bar{u}$ & $2\,\mathrm{MeV}$ & & \\
\hline 
Gluons & $g_i$ & $0\,\mathrm{MeV}$ & 1 & $8 \cdot 2$ \\
\hline 
Photons & $g_i$ & $0\,\mathrm{MeV}$ & 1 & $2$ \\
\hline 
\begin{tabular}{c}
Leptons
\end{tabular} 
& $\tau^{ \pm}$ & $1777\,\mathrm{MeV}$ & $\frac{1}{2}$ & $2$ \\
& $\mu^{ \pm}$ & $106\,\mathrm{MeV}$ & & \\
& $e^{ \pm}$ & $511\,\mathrm{keV}$ & & \\
\hline 
Neutrinos & $\nu_{\tau},\bar\nu_{\tau}$ & $0\,\mathrm{MeV}$ & $\frac{1}{2}$ & $1$ \\
& $\nu_{\mu},\bar\nu_{\mu}$ & $0\,\mathrm{MeV}$ & & \\
& $\nu_{e},\bar\nu_{e}$ & $0\,\mathrm{keV}$ & & \\
\hline 
Vector Bosons & $W^{\pm}$ & $80\,\mathrm{GeV}$ & $1$ & $3$ \\
& $Z$ & $91\,\mathrm{GeV}$ & & \\
\hline 
Higgs Boson & $H^0$ & $125\,\mathrm{GeV}$ & $0$ & $1$ \\
\hline
\end{tabular}

\caption{Elementary particles in the Standard Model with their mass, spin and number of degrees of freedom \cite{ParticleDataGroup:2024cfk}.}
\label{table:qgp}
\end{table}

\subsection{Equation of State}
To express all thermodynamic quantities in terms of chemical potentials and temperature, it is essential to adopt an appropriate equation of state (EoS). At sufficiently high temperatures, the quark-gluon plasma (QGP) can be approximated as an ideal quantum gas, where its constituents behave as non-interacting particles. As the temperature decreases, however, strong interaction effects among quarks and gluons become increasingly important, necessitating the use of an interacting EoS derived from lattice QCD simulations. In the low-temperature regime, where the QGP is fully hadronized, the system can be effectively described by the Hadron Resonance Gas (HRG) model.

In this work, we explore the phase diagram by initially modeling the QGP, photons, and leptons as ideal quantum gases. This is done under both the `2+1’ and `2+1+1’ flavor QGP scenarios. We then incorporate a lattice QCD equation of state with `2+1' flavors to include the effects of strong interaction, and further extend the analysis to the `2+1+1' flavor case. In all scenarios, our results are systematically compared to the HRG model, which serves as a reliable representation of the hadronic phase at low temperatures.

In the following sections, we explore each of these thermodynamic regimes by implementing the corresponding EoS into Eq. (\ref{bcons}) and Eq. (\ref{lepasym}).

\subsubsection{Free Ideal Quark Gluon Plasma}
At high temperatures, above $T\sim600-700$ MeV, we can expect that the QGP is approaching an ideal non-interacting quantum gas. All thermodynamic functions are then described by the proper Fermionic or Bosonic distribution as follows:
\begin{align}
&n_i=\frac{g_i}{2 \pi^2} \int_{m_i}^{+\infty}
\frac{ d \epsilon \epsilon \sqrt{\epsilon^2-m_i^2}}{\epsilon^{\frac{\left(\epsilon-\mu_i\right)}{k_B T}} \pm 1} \label{free dens}\\
& \rho_i=\frac{g_i}{2 \pi^2} \int_{m_i}^{+\infty} d \epsilon \frac{\epsilon^2 \sqrt{\epsilon^2-m_i^2}}{e^{\frac{\left(\epsilon-\mu_i\right)}{k_B T}} \pm 1} \\
& p_i=\frac{g_i}{6 \pi^2} \int_{m_i}^{+\infty} d \epsilon \frac{\left(\epsilon^2-m_i^2\right)^{3 / 2}}{e^{\frac{\left(\epsilon-\mu_i\right)}{k_B T}} \pm 1} \\
& s_i=\frac{\rho_i+p_i}{T}-\frac{\mu_i n_i}{T},
\end{align}
where $i$ stands for leptons, photons, quarks, and gluons, $n_i$ is the particle number density, $\rho_i$ is the internal energy, $p_i$ is the pressure, and $\mu_i$ is the chemical potential of each species; ``$+$" is for Fermions``$-$" is for Bosons. As a first approximation, we employ the ideal gas equation of state for all the particle species considered. This approach corresponds to the so-called free quark-gluon plasma model. Even when the lattice QCD equation of state is used to describe quarks and gluons, photons and leptons are still treated as free ideal quantum gases.
\\ For improved numerical stability and efficiency, we replaced the original integral expressions with the JEL approximation \cite{Johns:1996ht}. This formulation expresses the Fermi-Dirac integrals as a compact bivariate polynomial series. 

\subsubsection{Fourth Order Lattice QCD EoS for 2+1 Flavors}
To introduce the interactions between quarks and gluons we can rely on an equation of state based on lattice QCD simulations. These simulations allow the evaluation of thermodynamic functions only at vanishing chemical potentials due to the well-known Fermion sign problem \cite{deforcrand2010simulatingqcdfinitedensity}. As a result, calculations must be performed at zero chemical potentials, with finite values accessed through expansion techniques. In this work, we adopt the Taylor expansion method introduced in \cite{PhysRevC.100.064910}. The core idea is to evaluate the pressure at zero chemical potential and expand it up to fourth order in the baryon, electric charge, and strangeness (BQS) chemical potentials, as follows:
\begin{align}
\frac{p\left(T, \mu_B, \mu_Q, \mu_S\right)}{T^4}&=\\\sum_{i, j, k}^{i+j+k\leq 4}\frac{1}{i!j!k!} \chi_{i j k}^{B Q }&\left(\frac{\mu_B}{T}\right)^i\left(\frac{\mu_Q}{T}\right)^j\left(\frac{\mu_S}{T}\right)^k.
\end{align}
The susceptibilities are obtained at zero chemical potentials from the pressure as follows:
\begin{equation}
\chi_{i j k}^{B O S}=\left.\frac{\partial^{i+j+k}\left(p / T^4\right)}{\partial\left(\frac{\mu_B}{T}\right)^i \partial\left(\frac{\mu_Q}{T}\right)^j \partial\left(\frac{\mu_S}{T}\right)^k}\right|_{\mu_B, \mu_Q, \mu_S=0}.
\label{chisdef}
\end{equation}
Once the pressure is known, one can derive the entropy density, energy density, particle number densities, and the speed of sound:
\begin{equation}
\begin{aligned}
\frac{s}{T^3} & =\left.\frac{1}{T^3} \frac{\partial p}{\partial T}\right|_{\mu_i}, \quad \frac{\epsilon}{T^4}=\frac{s}{T^3}-\frac{p}{T^4}+\sum_i \frac{\mu_i}{T} \frac{n_i}{T^3} \\
\frac{n_i}{T^3} & =\left.\frac{1}{T^3} \frac{\partial p}{\partial \mu_i}\right|_{T, \mu_j}, \quad c_s^2=\left.\frac{\partial p}{\partial \epsilon}\right|_{n_i}+\left.\sum_i \frac{n_i}{\epsilon+p} \frac{\partial p}{\partial n_i}\right|_{\epsilon, n_j}.
\end{aligned}
\end{equation}

From this, the net baryon and electric charge densities can be expressed as:
\begin{equation}
    \frac{n_B}{T^3}=\left.\frac{1}{T^3} \frac{\partial p}{\partial \mu_B}\right|_{T, \mu_B} \quad
\frac{n_Q}{T^3}=\left.\frac{1}{T^3} \frac{\partial p}{\partial \mu_Q}\right|_{T, \mu_Q}.
\end{equation}
At this stage, in Eq. (\ref{qneutr}), leptons are still treated as free quantum particles obeying Fermi-Dirac statistics, while the QGP contribution is provided directly by the lattice QCD equation of state. The susceptibilities employed in this work are taken from \cite{Abuali:2025tbd, jahan_2025_15123623}. They were obtained by combining the lattice data with $2+1$ flavors for temperatures 135 MeV$<T<$220 MeV, HRG data at low temperatures and assuming that at $T\simeq800$ MeV they are $\sim10\%$ away from the Stefan-Boltzmann limit.
Within this framework, quarks and gluons interact strongly among themselves, while their interactions with leptons occur via the weak force under conditions of beta equilibrium. Electromagnetic interactions are neglected.

\subsubsection{Lattice QCD EoS for 2+1+1 Flavors}
As in the free ideal case, we aim to investigate how the inclusion of charm quarks affects the trajectories in the interacting scenario as well. The lattice QCD simulations in \cite{Abuali:2025tbd, jahan_2025_15123623} have been performed only with up, down, strange quarks, and gluons. Here we want to show how we combined them with susceptibilities that include the charm quark.\\
Introducing the charm quark, we are also introducing the charm conserved charge. The EoS will then look as follows:
\begin{align}
\frac{p\left(T, \mu_B, \mu_Q, \mu_S,\mu_C\right)}{T^4}&=\\\sum_{i, j, k, m}^{i+j+k+m\leq 4}\frac{1}{i!j!k!m!} \chi_{i j k m}^{B Q S C}&\left(\frac{\mu_B}{T}\right)^i\left(\frac{\mu_Q}{T}\right)^j\left(\frac{\mu_S}{T}\right)^k\left(\frac{\mu_C}{T}\right)^m.
\end{align}
where $\mu_C$ is the charm chemical potential. We define the charm density ($n_C$) as:
\begin{equation}
    n_C=n_{charm}.
\end{equation}

\begin{figure}[H]
    \centering
     \tikz[baseline] \draw[black, thick, solid] (0,0.1) -- (0.5,0.1); $2+1$ flavors, \tikz[baseline] \draw[red, thick, dashed] (0,0.1) -- (0.5,0.1); $2+1+1$ flavors
    \includegraphics[width=0.23\textwidth]{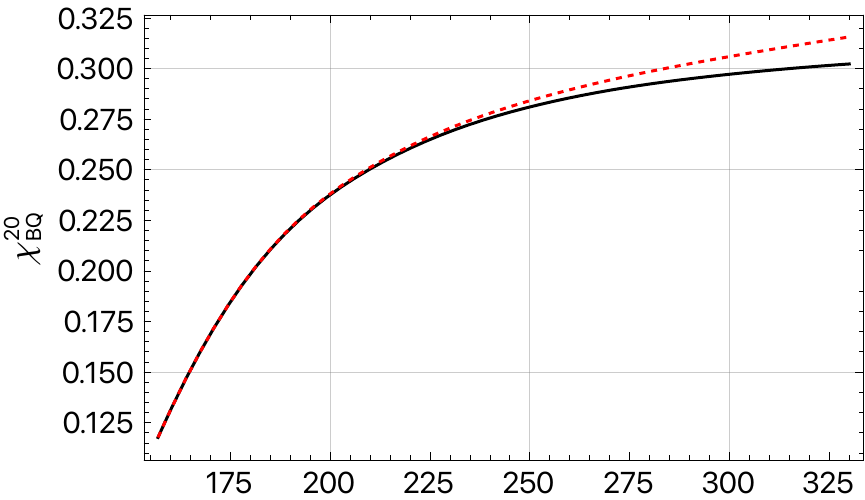}
    \includegraphics[width=0.23\textwidth]{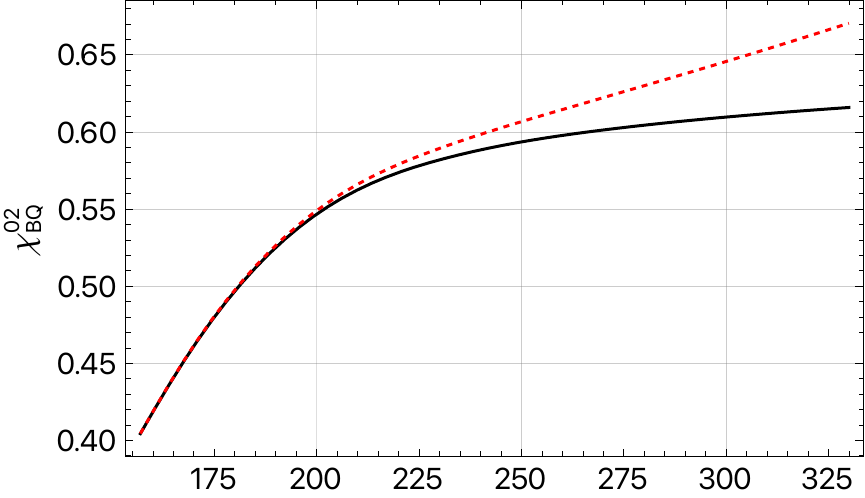}
    \includegraphics[width=0.23\textwidth]{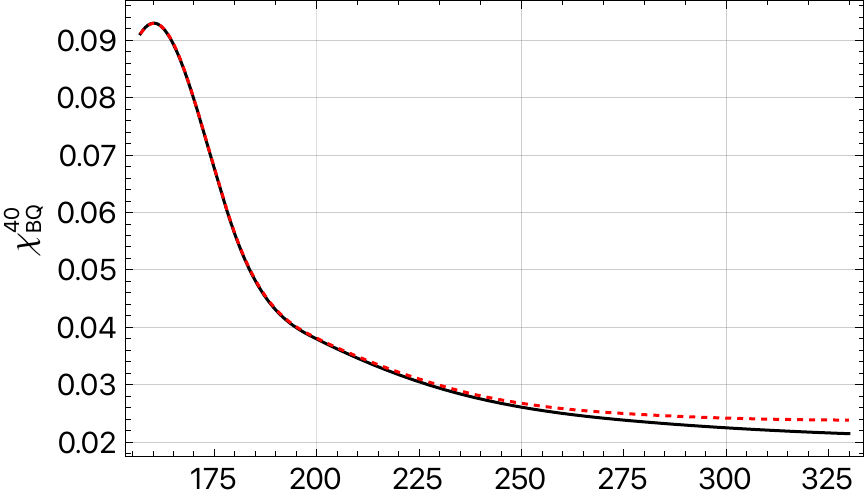}
    \includegraphics[width=0.23\textwidth]{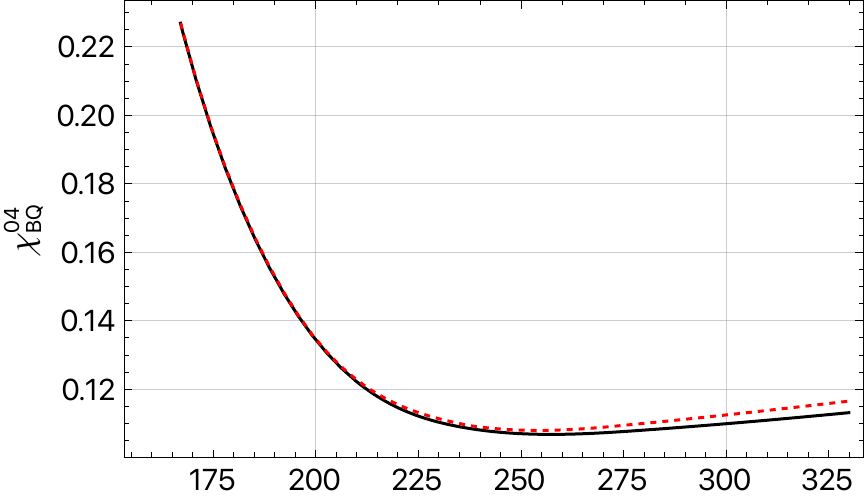}

    \vspace{0.1cm}

    \includegraphics[width=0.23\textwidth]{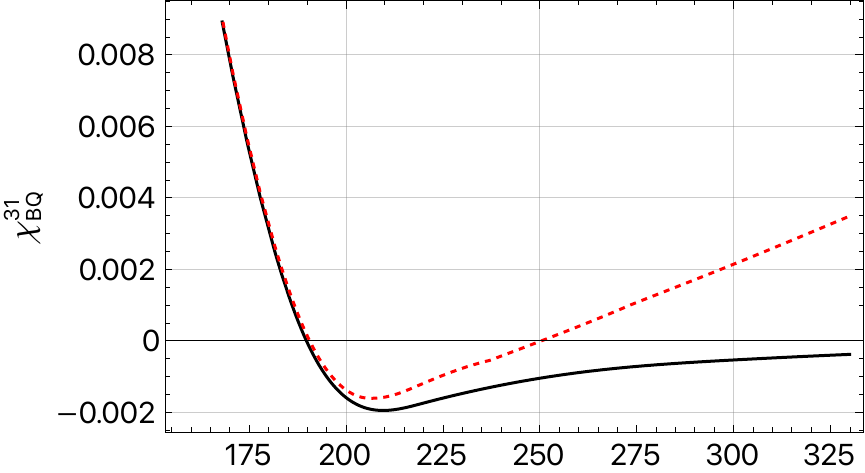}
    \includegraphics[width=0.23\textwidth]{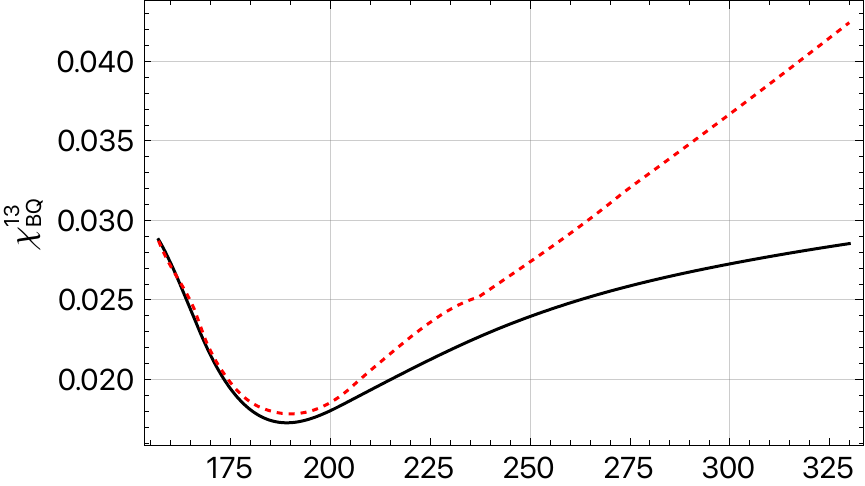}
    \includegraphics[width=0.23\textwidth]{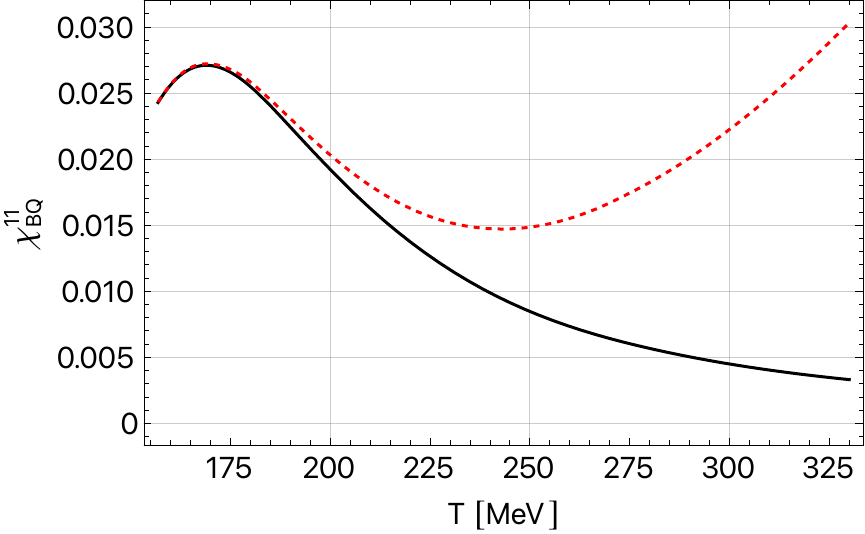}
    \includegraphics[width=0.23\textwidth]{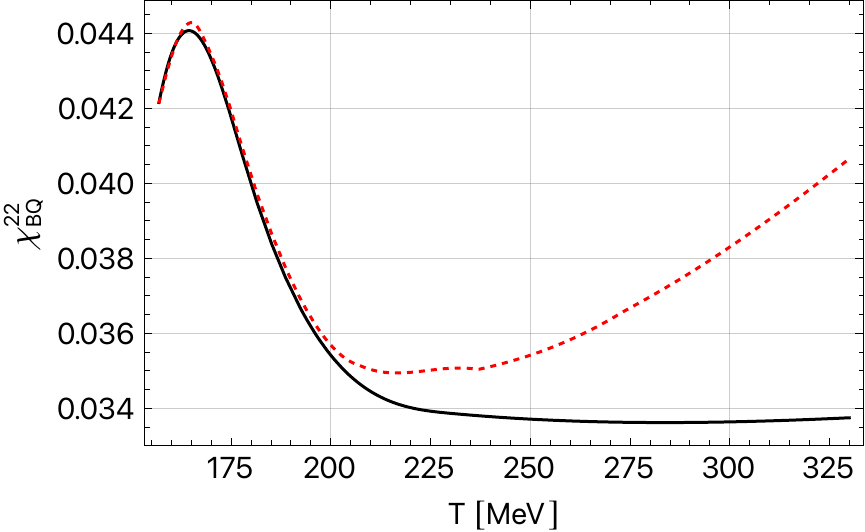}

    \caption{Susceptibilities in $B$ and $Q$ direction, with (red, dashed) and without (black, solid) the charm contribution.}
    \label{FigChis}
\end{figure}

Combining this and Eqs. (\ref{barion density})-(\ref{charge density}), we obtain:
\begin{equation}
    \mu_C=\mu_{up}-\mu_{charm}.
\end{equation}
Beta equilibrium still holds, which implies, from Eq. (\ref{beta equilibrium}), $\mu_C=0$ and $\mu_{charm}=\mu_{up}$. In essence, the only terms that remain in the EoS are those in the $B$ and $Q$ directions. Regardless of whether we directly consider the $S$ and $C$ directions, the susceptibilities in the $B$ and $Q$ directions are still influenced by the presence of the charm in the simulations.\\
From the definition of susceptibilities in Eq. (\ref{chisdef}), we can derive the relationship between them in the $ BQSC$ basis and $up \,\, down \,\, strange \,\, charm$ basis (in the calculation for the susceptibilities we use the formalism $u$, $d$, $s$ and $c$ instead of $up$, $down$, $strange$, $charm$ for a lighter formalism):
\begin{equation}
    \frac{d}{d\mu_B}=\sum_{i=u,d,s,c}\frac{\partial}{\partial \mu_{i}}\frac{\partial\mu_{i}}{\partial\mu_B}=\frac{1}{3}\partial_u+\frac{1}{3}\partial_d+\frac{1}{3}\partial_s+\frac{1}{3}\partial_c
\end{equation}
\begin{equation}
    \frac{d}{d\mu_Q}=\sum_{i=u,d,s,c}\frac{\partial}{\partial \mu_{i}}\frac{\partial\mu_{i}}{\partial\mu_Q}=\frac{2}{3}\partial_u-\frac{1}{3}\partial_d-\frac{1}{3}\partial_s+\frac{2}{3}\partial_c.
\end{equation}
Therefore, even if we look only at the $B$ and $Q$ directions, strange and charm quarks affect the EoS. Let us focus on $\chi_2^B$:
\begin{equation}
\begin{aligned}
   \frac{d^2}{d\mu_B^2} &=  \frac{1}{9} \left(\partial_u^2 + \partial_d^2 + \partial_s^2 + \partial_c^2 
    + 2\partial_u \partial_d + 2\partial_u \partial_s + 2\partial_u \partial_c \right. \\
    &\quad \left. + 2\partial_d \partial_s + 2\partial_d \partial_c + 2\partial_s \partial_c \right)
\end{aligned}
\end{equation}
The susceptibility becomes:
\begin{equation}
\begin{aligned}
   \chi_2^B = &\frac{1}{9} \left(\chi^2_u + \chi_d^2 + \chi_s^2 + \chi_c^2 
    + 2\chi^{11}_{ud} + 2\chi^{11}_{us} + 2\chi^{11}_{uc} \right. \\
    &\quad \left. + 2\chi^{11}_{ds} + 2\chi^{11}_{dc} + 2\chi^{11}_{sc} \right)
\end{aligned}
\end{equation}
The correction we introduce due to the charm is:
\begin{equation}
\begin{aligned}
   \chi_2^B =\chi_{2}^B|_{without charm}+ &\frac{1}{9} \left(\chi_c^2 
    + 2\chi^{11}_{uc} + 2\chi^{11}_{dc} + 2\chi^{11}_{sc} \right)
\end{aligned}
\end{equation}
In this work, we employ the EoS expanded up to fourth order in the chemical potentials $\mu_i$. Consequently, the susceptibilities of interest are $\chi_2^B$, $\chi_2^Q$, $\chi_{11}^{BQ}$, $\chi_4^B$, $\chi_4^Q$, $\chi_{31}^{BQ}$, $\chi_{13}^{BQ}$ and $\chi_{22}^{BQ}$. 
These susceptibilities have been computed excluding the charm-quark contribution. The charm data are from Ref. \cite{Bazavov_2024}. These data have been computed in the temperature range $156.8~\mathrm{\space MeV} \leq T \leq 330.2 ~\mathrm{\space MeV}$. \autoref{FigChis} shows the susceptibilities with and without the charm contribution. In all cases, the presence of the charm becomes effective for $T>200$ MeV. It is also evident that the effect is much more present in the off-diagonal terms $\chi_{BQ}^{11}$, $\chi_{BQ}^{22}$, $\chi_{BQ}^{13}$, $\chi_{BQ}^{31}$.

 \subsubsection{Hadron Resonance Gas Model}
 In the low-temperature regime, after the QCD transition, we have the hadronic phase, where most protons, neutrons, pions and other stable hadrons strongly interact with each other. A very simple and elegant way to describe this phase of the universe is through the HRG model \cite{PhysRevLett.118.182301,Hagedorn:1965st}. The idea is to describe the hadronic phase as composed of all hadrons and resonances as stable free particles described by the proper quantum distribution. The presence of resonances mimics the effect of the interaction. In this work we will verify that the Lattice QCD-based EoS converges to the HRG model for low temperatures. The trajectories for this phase have been calculated with the \texttt{Thermal-FIST} software developed by V. Vovchenko \cite{Vovchenko:2019pjl,VOVCHENKO2019295}, assuming interacting pions in order to have a more realistic description.

\subsection{Lepton asymmetries}
As long as the temperature does not drop below $ T_{\text{osc}} $, there are no processes capable of modifying the lepton asymmetries, as this would require interactions that violate lepton flavor and CP symmetry. These kind of processes start to be efficient at $ T\sim 10\,\text{MeV} $, when neutrinos start to oscillate and lepton flavor transports are allowed. Therefore, we are sure that in our temperature range the different $\ell_i $ can be considered constant.

In general, the constraints for $\ell_i $ come from the BBN and the CMB. Since $ \nu_e $ are directly involved in the beta reaction that establishes the ratio $ n_p / n_N $ (proton density over neutron density), while $ \nu_\mu $ and $ \nu_\tau $ are not, the BBN is directly sensitive only to $ \nu_e $, thus yielding strong constraints for $\ell_e $. Moreover, each lepton asymmetry introduces an extra contribution to the effective density $ n_{\text{eff}} $, which in turn changes the rate of expansion in the early Universe. This again can alter the abundances of light elements. This implies that, from the BBN, we can give some constraint on $\ell_e $ and on some combination of $\ell_e$, $\ell_\mu $ and $\ell_\tau $, but not on the three single lepton asymmetries. The CMB on the other hand is sensitive only to $ n_{\text{eff}} $, so from it we can obtain only some constraints to a combination of the three lepton asymmetries.

The values of $\ell_i $ across the QCD phase transition remain constant, but when we are at $ T < T_{\text{osc}} $ the three asymmetries change over time. The total lepton asymmetry $\ell =\ell_e +\ell_\mu +\ell_\tau $ always remains constant. The way in which the single lepton asymmetry evolves strictly depends on the value of $ \theta_{13} $ of the PMNS matrix, for which the main results are reported in \cite{mangano2011constraining,Castorina_2012, Mangano_2011, SergioPastor_2010}. What happens is that neutrino oscillations equalize the lepton asymmetries very rapidly. In particular, $\ell_\mu $ and $\ell_\tau $ can be assumed to be equal in absolute value, much earlier than the BBN, and independently of $ \theta_{13} $. This point is crucial: depending on the value of $ \theta_{13} $, $\ell_\tau $ and $\ell_\mu $ may or may not be equal to $\ell_e $. For example, in \cite{Mangano_2012}, Mangano et al. consider two situations: $ \sin^2(\theta_{13}) = 0 $ and $ \sin^2(\theta_{13}) = 0.04 $. In the first case, the lepton asymmetries do not equalize, while for the second case they do, highlighting the importance of the $ \theta_{13} $ dependence.

As anticipated, from the BBN we have the strongest constraint, which is $|\ell_e| \leq 10^{-3}$. This implies that, if $ \sin^2(\theta_{13}) = 0.04 $, the lepton asymmetries are equal to $\ell_e $ and we get that the total lepton asymmetry must be $\ell = \ell_e + \ell_\mu + \ell_\tau \leq 10^{-2} $. Otherwise, the lepton asymmetries do not equalize, and the constraints on $\ell_\mu $ and $\ell_\tau $ are quite loose, of order 1. Today, we measure $ \sin^2(\theta_{13}) = (2.20 \pm 0.07) \cdot 10^{-2} $ with a confidence level of $ 2\sigma $ \cite{ParticleDataGroup:2024cfk}; however, when considering the $ 3\sigma $ confidence level, the values previously discussed are still allowed. As discussed in \cite{PhysRevD.95.043506}, in the case of non-equalization of the lepton asymmetries, the BBN allows a total asymmetry $\ell \approx 1$, while in \cite{Oldengott_2017} they claim that the CMB imposes $\ell \leq 0.08$.

Since today the measure of $\theta_{13}$ at $3\sigma$ is quite loose, we do not know if the lepton asymmetries are equal or not, and we have two different upper limits to the total lepton asymmetry. The most stringent constraint we have is that $\ell_e \leq 10^{-3}$, and we will have to make the most of this condition.

From this brief discussion we conclude that, in general, we can assume $ |\ell_\mu| = |\ell_\tau| $, and the lepton asymmetry $\ell $ should be taken to satisfy $\ell \leq 10^{-2} $. Therefore, we consider two possible configurations for the lepton asymmetries:
\begin{itemize}
    \item Symmetric case: \begin{equation}
    \ell_e = \ell_\tau = \ell_\mu= \frac{\ell}{3} 
    \label{eq:config1}
\end{equation}
\item  Asymmetric case: \begin{align}
    \ell_e &= 0 \label{eq:config2} \\
    \ell_\mu &= -\ell_\tau . \label{eq:config3}
\end{align}
\end{itemize}
These two configurations are the most discussed in the literature, in particular in \cite{gao2022cosmology,PhysRevLett.121.201302,PhysRevD.105.123533}. In general, we are going to check how the value of $\ell_i$ changes the cosmic trajectories and how much we have to increase them, in order to reach the range of chemical potentials corresponding to the most recent critical point predictions. In fact, as we will see, by increasing their value, the cosmological trajectories move to larger values of the chemical potentials.

\section{Results}
\subsubsection{Cosmic Trajectories}

\begin{figure}[h]
    \textbf{Symmetric case; free quantum gas}
    \includegraphics[width=1\linewidth]{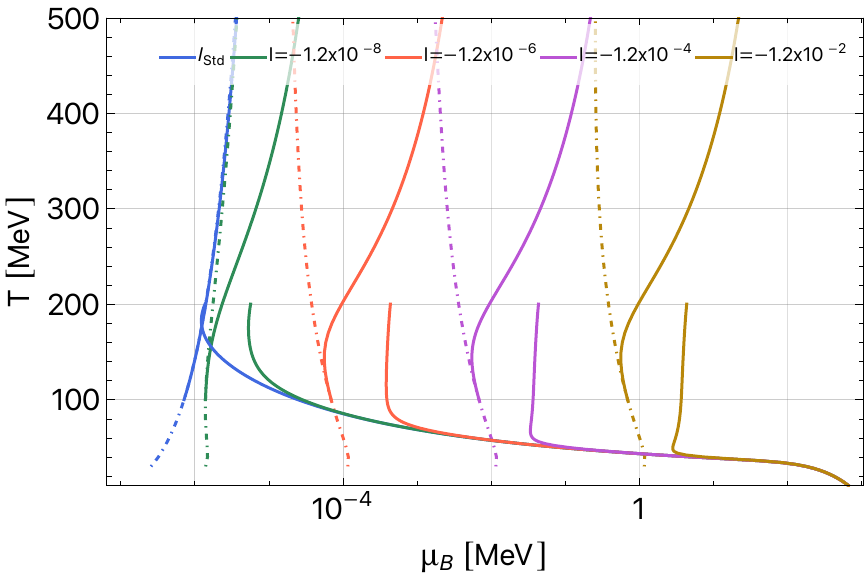}    \makeatletter\long\def\@ifdim#1#2#3{#2}\makeatother
    \caption{Cosmic trajectories computed for the symmetric lepton‐asymmetry configuration. The lower solid curves denote the HRG model, the upper solid curves correspond to $2+1+1$ flavor QGP trajectories, and the dot‐dashed curves represent $2+1$ flavor QGP trajectories.}
    \label{freeqgpsym}
\end{figure}

\begin{figure}[]
    \centering
    \textbf{Symmetric lepton asymmetries \\ Three flavors}
    \includegraphics[width=1\linewidth]{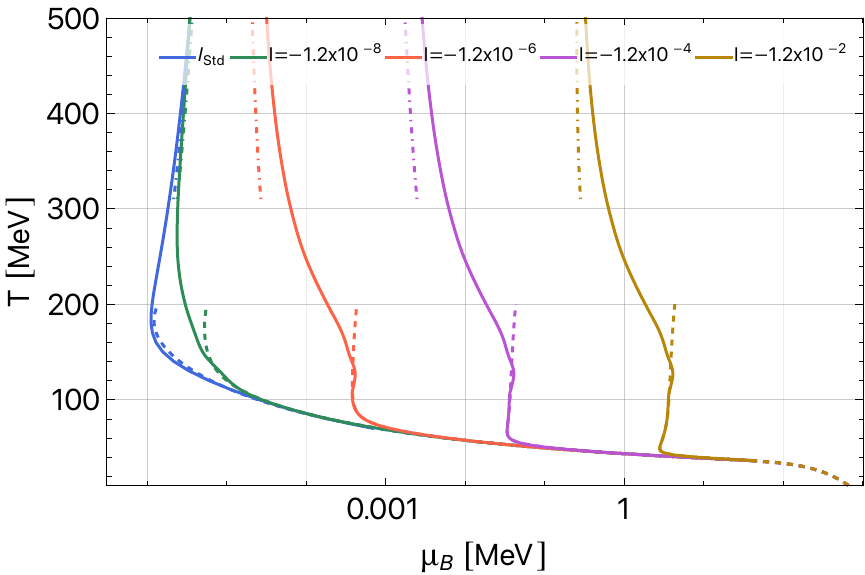}
    \includegraphics[width=1\linewidth]{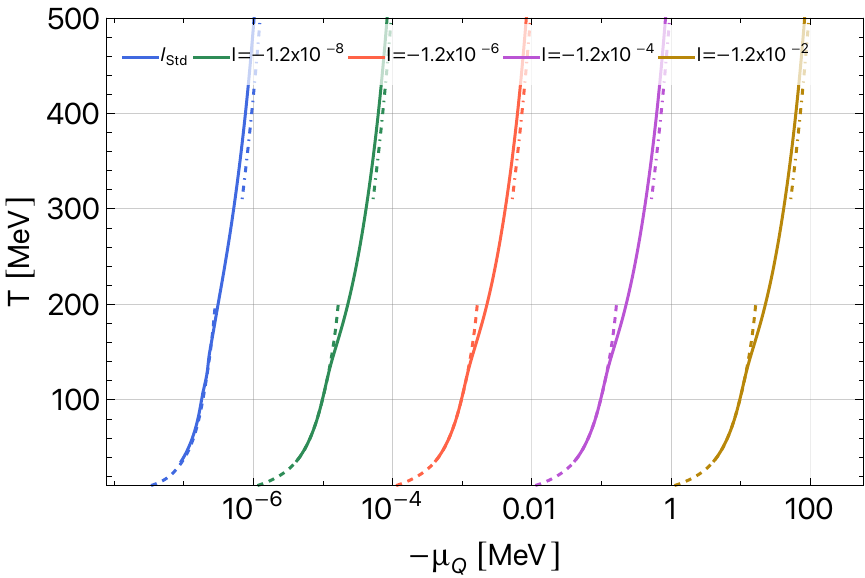}
    \caption{Cosmic trajectories calculated for a $2+1$ flavor QGP in the case of symmetric lepton asymmetries in the $(T,~\mu_B)$ (top panel) and $(T,~-\mu_Q)$ (bottom panel) planes. Solid lines show the results obtained with the lattice QCD EoS; the dashed lines are the result from the HRG model; the dot-dashed lines represent the free QGP results.}
    \label{threeflavint}
\end{figure}

\begin{figure}[]
    \centering
    \textbf{Symmetric lepton asymmetries \\ Four flavors}
    \includegraphics[width=1\linewidth]{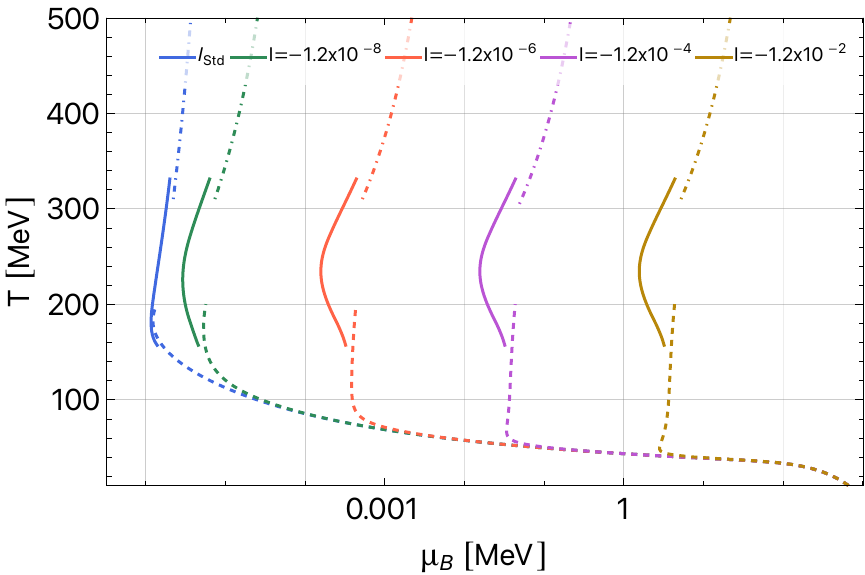}
    \includegraphics[width=1\linewidth]{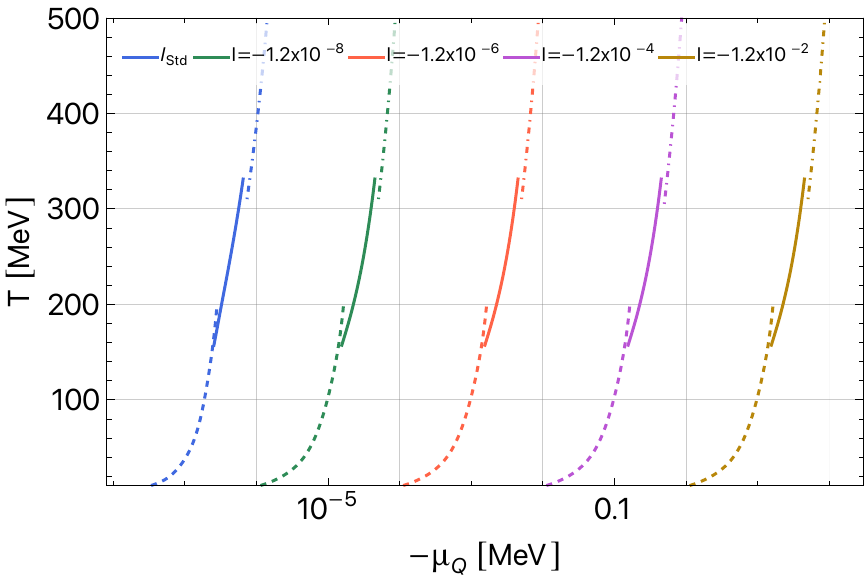}
    \caption{Cosmic trajectories calculated for a $2+1+1$ flavor QGP in the case of symmetric lepton asymmetries in the $(T,~\mu_B)$ (top panel) and $(T,~-\mu_Q)$ (bottom panel) planes. Solid lines show the results obtained with the lattice QCD EoS; the dashed lines are the result from the HRG model; the dot-dashed lines represent the free QGP results. Notice that the lattice QCD trajectories do not extend below $T\sim150$ MeV and above $T\sim330$ MeV, due to the limited range of the lattice QCD results for the charm quark susceptibilities.}
    \label{fourflavint}
\end{figure}

\begin{figure}[]
    \centering
    \textbf{Asymmetric lepton asymmetries \\ Four flavors}
    
    \includegraphics[width=1\linewidth]{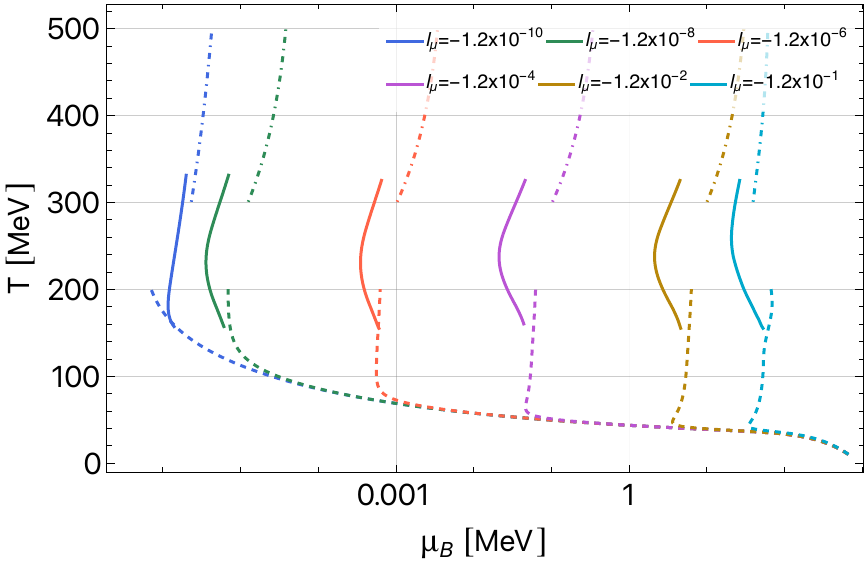}
    \includegraphics[width=1\linewidth]{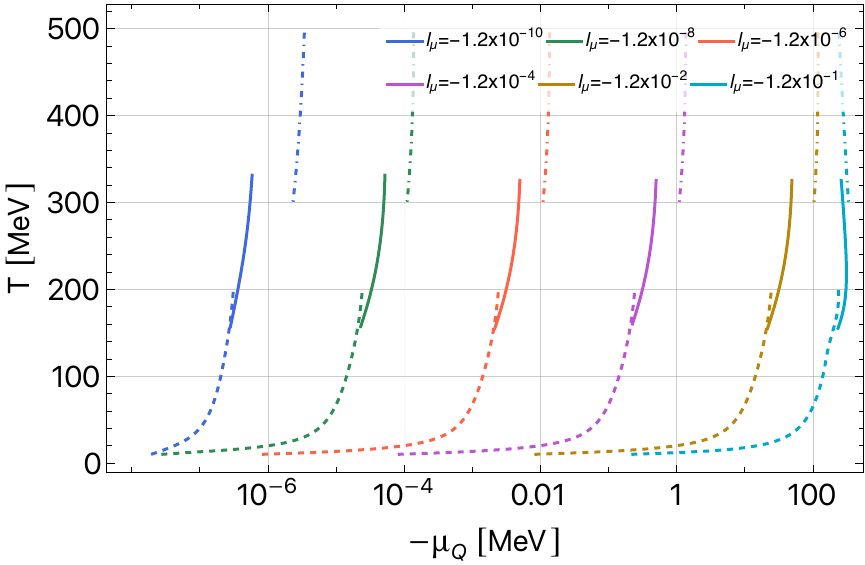}
    \caption{Cosmic trajectories calculated for a $2+1+1$ flavor QGP in the case of asymmetric lepton asymmetries in the $(T,~\mu_B)$ (top panel) and $(T,~-\mu_Q)$ (bottom panel) planes. Solid lines show the results obtained with the lattice QCD EoS; the dashed lines are the result from the HRG model; the dot-dashed lines represent the free QGP results. Notice that the lattice QCD trajectories do not extend below $T\sim150$ MeV and above $T\sim330$ MeV, due to the limited range of the lattice QCD results for the charm quark susceptibilities. }
    \label{foureasym}
\end{figure}

In order to understand the nature of the cosmic QCD phase transition, we compare the obtained trajectories with the (expected) critical end point. The most recent estimate locates the CEP at $\left(T_c, \mu_{B, c}\right) = (114.3 \pm 6.9,\ 602.1 \pm 62.1)\ \mathrm{MeV}$ \cite{Shah:2024img}, albeit at $\mu_Q=0$. Our trajectories have been calculated for the following values of $\ell$: the standard case $\ell = -\frac{51}{28} \cdot b = \ell_{\mathrm{std}}$, and the additional values $\ell = -1.2 \times 10^{-8}$, $\ell = -1.2 \times 10^{-6}$, $\ell = -1.2 \times 10^{-4}$, and $\ell = -1.2 \times 10^{-2}$, for the symmetric case. For the asymmetric case, where $\ell_e=0, \space~ \ell_\mu=-\ell_\tau$, we calculated the trajectories for  $\ell_\mu = -1.2 \times 10^{-8}$, $\ell_\mu = 1.2 \times 10^{-6}$, $\ell_\mu = -1.2 \times 10^{-4}$, $\ell_\mu = -1.2 \times 10^{-2}$ and $\ell_\mu = -1.2 \times 10^{-1}$, since with this configuration larger values of $\ell_\mu$ are allowed. \\
In \autoref{freeqgpsym} we show the cosmic trajectories obtained using the free QGP model for both $2+1$ flavors (solid lines) and $2+1+1$ flavors (dot-dashed lines). These trajectories converge to each other for $T \sim 150 ~\mathrm{MeV}$, but never approach the HRG model. In contrast, when the lattice QCD equation of state is used (\autoref{threeflavint}, \autoref{fourflavint} and \autoref{foureasym}), the trajectories consistently converge to the HRG model at low temperatures, confirming the well-behaved nature of the EoS at low temperatures. The $2+1$ and $2+1+1$ flavor trajectories calculated using the lattice EoS also converge to each other at $T \sim 150~\mathrm{MeV}$. Since we verified this in the symmetric case, for the asymmetric case we only show the four flavor calculation in \autoref{foureasym}. From  \autoref{threeflavint} and \autoref{fourflavint} we also confirm that, at high temperatures, the results obtained with the lattice QCD EoS converge to the ideal gas EoS results. When the largest value of the muon lepton asymmetry is considered, the hadron gas trajectory becomes discontinuous around $T \sim 100\ \text{MeV}$ due to pion excitations (\autoref{foureasym}, $\ell_\mu=-1.2\cdot 10^{-1}$).\\

\begin{figure}[h]
    \textbf{3D representation of the Asymmetric case }
    \includegraphics[width=1\linewidth]{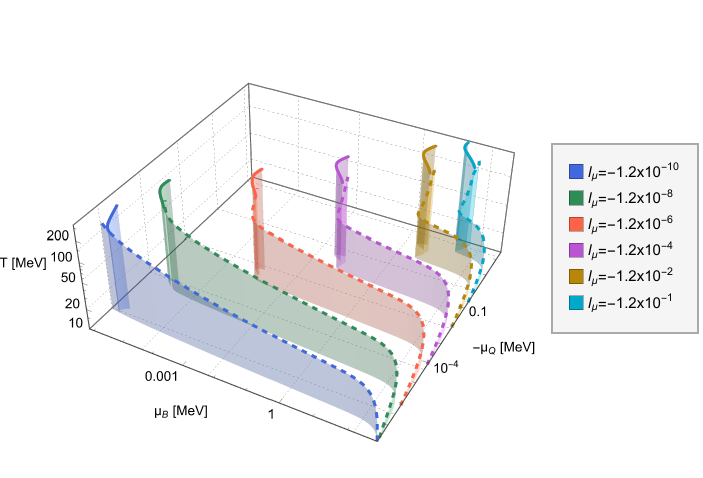}    \makeatletter\long\def\@ifdim#1#2#3{#2}\makeatother
    \caption{Cosmic trajectories calculated for the asymmetric configuration of the lepton asymmetries, represented in the 3D ($\mu_B, -\mu_Q,T$) space. Dashed lines represent the HRG model results while solid lines represent the $2+1+1$ flavor lattice QCD EoS.}
    \label{3DTraj}
\end{figure}

In all scenarios considered, the trajectories never reach large enough values of the chemical potentials, remaining to the left of the critical point proposed in \cite{Shah:2024img}, even when the largest value of $\ell_\mu$ is considered in the asymmetric case. This implies that the QCD phase transition is most likely a crossover, instead of a first order phase transition. It is important to note that the proposed critical point has been computed under the assumption of zero electric charge chemical potential, which does not apply to our case. In our setup, the electric charge chemical potential is of the same order of magnitude as the baryon chemical potential. One might expect that introducing a non-vanishing $\mu_Q$ would move the critical point to smaller values of $\mu_B$, but further investigations are needed for a conclusive statement to be made. \\
In \autoref{3DTraj} we show a 3D representation of the cosmic trajectories in the asymmetric case, calculated with the HRG model and the $2+1+1$ flavor lattice QCD EoS.

\subsubsection{Entropy density}
\begin{figure}[]
    \centering
    \textbf{Entropy density evolution}\\
    Symmetric case
    \includegraphics[width=1\linewidth]{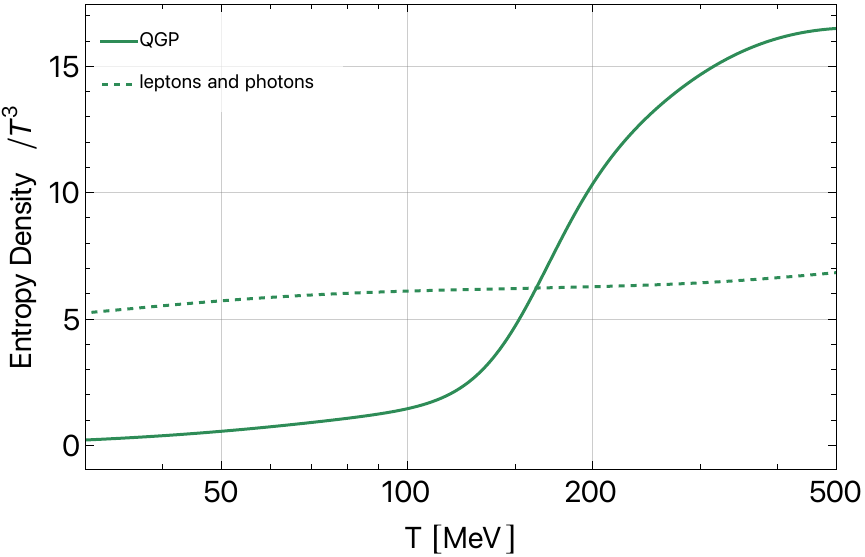}
    Asymmetric case
      \includegraphics[width=1\linewidth]{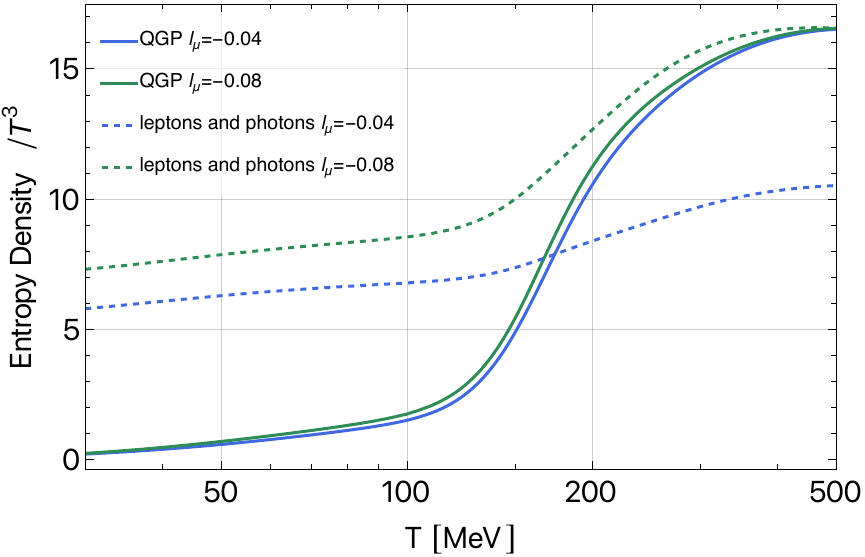}
    \caption{Entropy density evolution computed with the $2+1$ flavor lattice QCD equation of state. Solid curves trace the QGP entropy density, while dashed curves show the lepton+photon entropy density. \textit{Top panel}: symmetric scenario, where all lepton‐asymmetry values overlap. \textit{Bottom panel}: asymmetric scenario.
    } 
    \label{fig:entropydensevo}
\end{figure}

Another useful application of computing the cosmic trajectories is the ability to track the evolution of thermodynamical quantities. In \autoref{fig:entropydensevo}, we show the evolution of the entropy density for both the symmetric and asymmetric cases. In the symmetric scenario, the evolution is independent of the lepton asymmetry, whereas in the asymmetric case it explicitly depends on it (we show the entropy evolution for $\ell_\mu=-0.04$ and $\ell_\mu=-0.08$). The difference between these two regimes lies in the lepton abundances: in the symmetric case, the populations of electrons, muons, and tauons are always equal, while in the asymmetric case, the electron asymmetry is set to zero. Because weak beta processes are predominantly mediated by electrons, they remain the principal mediators in the symmetric case. However, in the asymmetric case, the lack of electrons reduces their contribution to the beta processes, shifting the mediating role to the muons. As muons are less efficient in mediating beta interactions and the entropy is sensitive to their abundance, the evolution of the entropy becomes dependent on the muon asymmetry. Increasing $\ell_\mu$ increases the number of available muons, so the bigger $\ell_\mu$, the more they dominate beta interactions. This implies the increase of the system's entropy. \\
We also observe that the transition from the QGP to the hadron gas is clearly visible in the temperature range $\sim120\,\mathrm{MeV} < T < 300\,\mathrm{MeV}$, where the equation of state is obtained from lattice simulations. Furthermore, the evolution of the entropy density exhibits a strong temperature dependence in the QGP sector, while it remains comparatively weaker for leptons and photons.
\section{Conclusion}
In this work, we computed the $5+1$ dimensional cosmic trajectories followed by the early Universe during the QCD phase transition as functions of the lepton asymmetries, treated as free parameters of the system. 
We employed the state-of-the-art lattice QCD equation of state based on a fourth-order Taylor expansion in the chemical potentials, as proposed in \cite{PhysRevC.100.064910}, incorporating the most recent $2+1$ flavor susceptibilities from \cite{jahan_2025_15123623}. We also combined these susceptibilities with those that include the charm quark from Ref. \cite{Bazavov_2024}, thereby employing a $2+1+1$ flavor lattice QCD EoS expanded up to fourth-order in the chemical potentials. We have also shown how the entropy density evolution of the Universe is affected by different configurations of the lepton asymmetries. \\

We investigated two distinct configurations of the lepton asymmetries: the symmetric case ($\ell_e = \ell_\mu = \ell_\tau = \ell/3$) and the asymmetric case ($\ell_e = 0$, $\ell_\mu = -l_\tau$, with $\ell = 0$). In both scenarios, we find that increasing either the total lepton asymmetry $\ell$ or  $\ell_\mu$ shifts the cosmic trajectories toward larger values of the chemical potentials, confirming the behavior proposed in \cite{wygas2018cosmic, PhysRevD.105.123533}. However, the chemical potentials reached remain insufficient to trigger a first-order QCD phase transition, confirming the conclusion that the transition is most likely a crossover, even when the most exotic configuration with large $\ell_\mu$ and zero $\ell_e$ is considered. 
The asymmetric case generally results in a higher entropy density in the system. This is because, when $\ell_e = 0$ and $\ell_\mu \neq 0$, the beta-processes are predominantly mediated by muons, for which the larger mass makes these interactions less efficient, increasing the entropy. \\

Analyzing the trajectories obtained using the lattice QCD equation of state, we confirm its consistent behavior across the entire temperature range: the trajectories approach the Hadron Resonance Gas (HRG) model at low temperatures and converge to the Stefan–Boltzmann limit at high temperatures. Moreover, the inclusion of the charm quark does not significantly alter the nature of the transition, as its contribution becomes relevant only at temperatures around $T \simeq 200~\text{MeV}$. Nevertheless, in the high-temperature regime, the presence of the charm quark must be taken into account, as it sizeably modifies the cosmic trajectories.

Several studies \cite{Fu:2019hdw,Gunkel:2021oya,Gao:2020fbl,Hippert:2023bel,Basar:2023nkp,Clarke:2024ugt,Shah:2024img} indicate that the critical endpoint likely lies at large baryon chemical potentials ($\mu_Q=0$). Our findings show that a first-order phase transition is inaccessible within the conventional scenario characterized by constrained total lepton asymmetry. Specifically, in the asymmetric configuration (with $\ell_e = 0$ and varying $\ell_\tau$), the baryon chemical potential $\mu_B$ remains bounded at fixed $T$ (see \autoref{bouncingtra}), precluding unbounded growth even for large $\ell_\tau$. Thus, in such a constrained scenario, the cosmic trajectory cannot enter a first-order phase transition region unless an additional, distinct first-order transition, such as pion condensation, is introduced \cite{Ferreira:2025zeu}. In our approach, we account for interacting pions using the \texttt{Thermal‐FIST} code. This yields a second‐order phase transition \cite{Vovchenko:2020crk} although, depending on the model considered, we cannot exclude the possibility of a first‐order transition associated with pion condensation at high temperatures.

\begin{acknowledgements}
We thank P. Petreczky for providing the lattice QCD results for the quark number susceptibilities including the charm quark. We thank M. Lattanzi and V. Vovchenko for fruitful discussions and constructive comments.
This material is based upon work supported by the National Science Foundation under grants No. PHY-2208724, PHY-2116686 and PHY-2514763, and within the framework of the MUSES collaboration, under Grant No. OAC-2103680. This material is also based upon work supported by the U.S. Department of Energy, Office of Science, Office of Nuclear Physics, under Award Number DE-SC0022023, as well as by the National Aeronautics and Space Agency (NASA) under Award Number 80NSSC24K0767.       
\end{acknowledgements}
\appendix
\section{Analytical behavior }
\subsection{Symmetric case: linear behavior}
\begin{figure}[H]
    \centering
    \textbf{Symmetric case: $\ell=-1.2\cdot 10^{-1}$}
\includegraphics[width=1\linewidth]{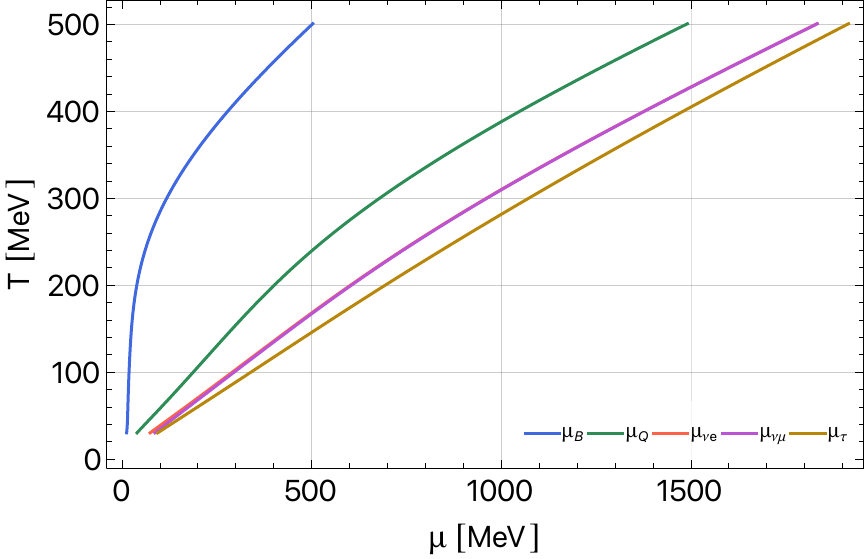}
    \caption{Cosmic trajectory for different chemical potentials (module) and $\ell=-1.2\cdot 10^{-1}$ in the free QGP case. Notice that the chemical potentials for electron and muon neutrinos are identical.}
    \label{sym all  pot}
\end{figure}

In this paragraph, we want to verify how the trajectories depend on the lepton asymmetries. Firstly, we start from the symmetric configuration, and then we will check the asymmetric case.
We want to understand how the chemical potentials are related to the lepton asymmetries when $\ell$ increases ($\ell\gg b$ case). To simplify the analysis, we assume that all particles are massless:
\begin{equation}
n = \frac{\partial p}{\partial \mu} = g \left( \frac{2}{21} T^2 \mu + \frac{2}{21 \pi^2} \mu^3 \right)
\end{equation}
\begin{equation}
s = g \left( \frac{2 \pi^2}{45} T^3 + \frac{2}{21} \mu^2 T \right),
\end{equation}
where $n$ is the net density of the particle number, $s$ is the entropy density, $\mu$ is the chemical potential and $T$ the temperature, $g$ are the degrees of freedom. 
In \autoref{sym all  pot}, we show the chemical potentials for the symmetric configuration with $\ell=-1.2\cdot 10^{-2}$. From this plot, we can observe that for $\ell\gg b$ the neutrino chemical potentials and electric charge chemical potential are much larger than the baryo-chemical potential. From here we assume that, in this limit, the baryo-chemical potential can be neglected. Since we are also assuming all particles to be massless, the three leptons are indistinguishable: $\mu_e=\mu_{\mu}=\mu_{\tau}=\mu_l$, then also $\mu_{\nu e}=\mu_{\nu \mu}=\mu_{\nu \tau}=\mu_{l\nu}$. Combining this with Eq. (\ref{qneutr}) and Eq. (\ref{lepton chem}), we obtain:
\begin{align}
    \mu_l&\approx 0.42\, \mu_Q\\
    \mu_{l\nu}&\approx 1.42\, \mu_Q.
\end{align}
Now, combining this with Eq. (\ref{lepasym}) and Eq. (\ref{bcons}), we finally obtain the equations of interest:
\begin{equation}
    \frac{\mu_Q}{T}\approx25 \,\ell
\end{equation}
\begin{equation}
    \frac{\mu_B}{T}\approx 65 \,b-12.5 \,\ell.
\end{equation}
These equations are rough approximations but are quite significant: $\ell$ will increase in the negative direction, this means that increasing it we can expect the trajectories to move to larger values of the baryon chemical potential and to larger and negative charge chemical potential. This implies that, by increasing $\ell$, we get closer to the possible critical point.
\label{l sym analytical}
\subsection{Asymmetric case: bouncing trajectories}\label{bouncingtra}
In the analysis of cosmic trajectories characterized by asymmetric lepton asymmetries, specifically the case defined by
\begin{equation}
\ell_e = 0,\quad
\ell_\mu = -\ell_\tau,\quad
\ell_\tau > 0,
\end{equation}
and a standard baryon asymmetry $b$, a distinctive behavior emerges in the evolution of the baryon chemical potential $\mu_{B}$. The trajectories exhibit a pronounced maximum in $\mu_{B}$, the location of which depends on the chosen value of $\ell_{\tau}$.

In this asymmetric case, we do not present an analytic approximation for $\mu_\mathrm{B}(T)$, since in the bounce region one finds $\mu_\mathrm{B} \sim T$, not allowing for a controlled expansion.

This counterintuitive behavior, contrary to that in the symmetric case, can be understood by considering the relationship between the $\tau$‐lepton chemical potential, $\mu_{\tau}$, and the $\tau$ mass, $m_{\tau}$.

Indeed, the maxima in $\mu_\mathrm{B}$ occur precisely when
\begin{equation}
\mu_\tau = m_\tau \,.
\end{equation}
Below this threshold, $\tau$ leptons are Boltzmann-suppressed and contribute negligibly to the total charge density. As soon as $\mu_\tau$ exceeds $m_\tau$, a nonrelativistic degenerate $\tau$ Fermi sea develops and the $\tau$ number density rises sharply. This sudden increase in the $\tau$ population rearranges the chemical equilibrium conditions among quarks and leptons, thereby producing the observed extremum in $\mu_\mathrm{B}$.

\begin{figure}[H]
\centering
\includegraphics[width=.5\textwidth]{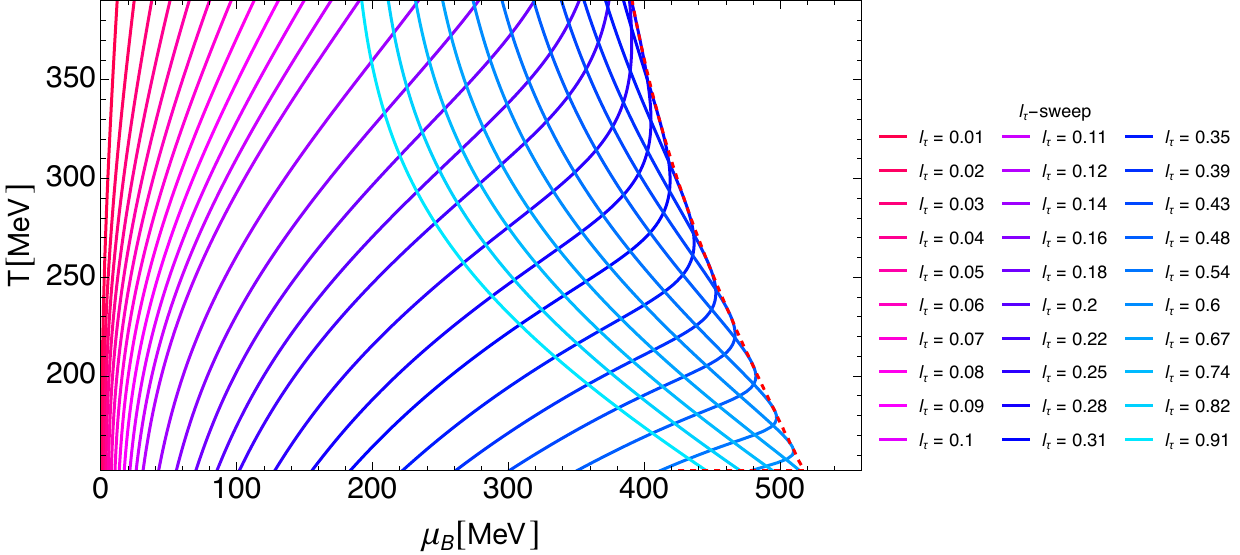}
    \caption{Cosmological trajectories for a free-QGP in the asymmetric lepton-asymmetry case ($\ell_e \;\simeq\;0$, $\ell_\mu \;=\;-\,\ell_\tau \;\neq\;0$) with standard baryon asymmetry. Solid lines indicate trajectories corresponding to the values of $\ell_\tau$ shown in the legend; the dashed red line represents the limiting values of the trajectories, obtained by imposing the condition $\mu_\tau = m_\tau$. As the magnitude of the $\tau$-lepton asymmetry grows beyond a critical threshold, each trajectory ``bounces" and turns toward lower baryon chemical potentials.}
    \label{bounce}
\end{figure}

This interpretation is consistent across the range of values of $\ell_\tau$ analyzed. As visible in \autoref{bounce} the limiting trajectory in the $T$-$\mu_\mathrm{B}$ plane is the one obtained by imposing $\mu_\tau = m_\tau$. 

\newpage

\bibliographystyle{apsrev4-2}
\bibliography{main}

\begin{thebibliography}{55}%
\makeatletter
\providecommand \@ifxundefined [1]{%
 \@ifx{#1\undefined}
}%
\providecommand \@ifnum [1]{%
 \ifnum #1\expandafter \@firstoftwo
 \else \expandafter \@secondoftwo
 \fi
}%
\providecommand \@ifx [1]{%
 \ifx #1\expandafter \@firstoftwo
 \else \expandafter \@secondoftwo
 \fi
}%
\providecommand \natexlab [1]{#1}%
\providecommand \enquote  [1]{``#1''}%
\providecommand \bibnamefont  [1]{#1}%
\providecommand \bibfnamefont [1]{#1}%
\providecommand \citenamefont [1]{#1}%
\providecommand \href@noop [0]{\@secondoftwo}%
\providecommand \href [0]{\begingroup \@sanitize@url \@href}%
\providecommand \@href[1]{\@@startlink{#1}\@@href}%
\providecommand \@@href[1]{\endgroup#1\@@endlink}%
\providecommand \@sanitize@url [0]{\catcode `\\12\catcode `\$12\catcode `\&12\catcode `\#12\catcode `\^12\catcode `\_12\catcode `\%12\relax}%
\providecommand \@@startlink[1]{}%
\providecommand \@@endlink[0]{}%
\providecommand \url  [0]{\begingroup\@sanitize@url \@url }%
\providecommand \@url [1]{\endgroup\@href {#1}{\urlprefix }}%
\providecommand \urlprefix  [0]{URL }%
\providecommand \Eprint [0]{\href }%
\providecommand \doibase [0]{https://doi.org/}%
\providecommand \selectlanguage [0]{\@gobble}%
\providecommand \bibinfo  [0]{\@secondoftwo}%
\providecommand \bibfield  [0]{\@secondoftwo}%
\providecommand \translation [1]{[#1]}%
\providecommand \BibitemOpen [0]{}%
\providecommand \bibitemStop [0]{}%
\providecommand \bibitemNoStop [0]{.\EOS\space}%
\providecommand \EOS [0]{\spacefactor3000\relax}%
\providecommand \BibitemShut  [1]{\csname bibitem#1\endcsname}%
\let\auto@bib@innerbib\@empty
%</preamble>
\bibitem [{\citenamefont {Espinosa}\ and\ \citenamefont {Quiros}(1993)}]{Espinosa:1993bs}%
  \BibitemOpen
  \bibfield  {author} {\bibinfo {author} {\bibfnamefont {J.~R.}\ \bibnamefont {Espinosa}}\ and\ \bibinfo {author} {\bibfnamefont {M.}~\bibnamefont {Quiros}},\ }\href {https://doi.org/10.1016/0370-2693(93)91111-Y} {\bibfield  {journal} {\bibinfo  {journal} {Phys. Lett. B}\ }\textbf {\bibinfo {volume} {305}},\ \bibinfo {pages} {98} (\bibinfo {year} {1993})},\ \Eprint {https://arxiv.org/abs/hep-ph/9301285} {arXiv:hep-ph/9301285} \BibitemShut {NoStop}%
\bibitem [{\citenamefont {Iso}\ \emph {et~al.}(2017)\citenamefont {Iso}, \citenamefont {Serpico},\ and\ \citenamefont {Shimada}}]{Iso:2017uuu}%
  \BibitemOpen
  \bibfield  {author} {\bibinfo {author} {\bibfnamefont {S.}~\bibnamefont {Iso}}, \bibinfo {author} {\bibfnamefont {P.~D.}\ \bibnamefont {Serpico}},\ and\ \bibinfo {author} {\bibfnamefont {K.}~\bibnamefont {Shimada}},\ }\href {https://doi.org/10.1103/PhysRevLett.119.141301} {\bibfield  {journal} {\bibinfo  {journal} {Phys. Rev. Lett.}\ }\textbf {\bibinfo {volume} {119}},\ \bibinfo {pages} {141301} (\bibinfo {year} {2017})},\ \Eprint {https://arxiv.org/abs/1704.04955} {arXiv:1704.04955 [hep-ph]} \BibitemShut {NoStop}%
\bibitem [{\citenamefont {Hambye}\ \emph {et~al.}(2018)\citenamefont {Hambye}, \citenamefont {Strumia},\ and\ \citenamefont {Teresi}}]{Hambye:2018qjv}%
  \BibitemOpen
  \bibfield  {author} {\bibinfo {author} {\bibfnamefont {T.}~\bibnamefont {Hambye}}, \bibinfo {author} {\bibfnamefont {A.}~\bibnamefont {Strumia}},\ and\ \bibinfo {author} {\bibfnamefont {D.}~\bibnamefont {Teresi}},\ }\href {https://doi.org/10.1007/JHEP08(2018)188} {\bibfield  {journal} {\bibinfo  {journal} {JHEP}\ }\textbf {\bibinfo {volume} {08}},\ \bibinfo {pages} {188}},\ \Eprint {https://arxiv.org/abs/1805.01473} {arXiv:1805.01473 [hep-ph]} \BibitemShut {NoStop}%
\bibitem [{\citenamefont {Schwarz}(2003)}]{Schwarz:2003du}%
  \BibitemOpen
  \bibfield  {author} {\bibinfo {author} {\bibfnamefont {D.~J.}\ \bibnamefont {Schwarz}},\ }\href {https://doi.org/10.1002/andp.200310010} {\bibfield  {journal} {\bibinfo  {journal} {Annalen Phys.}\ }\textbf {\bibinfo {volume} {12}},\ \bibinfo {pages} {220} (\bibinfo {year} {2003})},\ \Eprint {https://arxiv.org/abs/astro-ph/0303574} {arXiv:astro-ph/0303574} \BibitemShut {NoStop}%
\bibitem [{\citenamefont {Witten}(1984)}]{PhysRevD.30.272}%
  \BibitemOpen
  \bibfield  {author} {\bibinfo {author} {\bibfnamefont {E.}~\bibnamefont {Witten}},\ }\href {https://doi.org/10.1103/PhysRevD.30.272} {\bibfield  {journal} {\bibinfo  {journal} {Phys. Rev. D}\ }\textbf {\bibinfo {volume} {30}},\ \bibinfo {pages} {272} (\bibinfo {year} {1984})}\BibitemShut {NoStop}%
\bibitem [{\citenamefont {Jedamzik}(1997)}]{Jedamzik:1996mr}%
  \BibitemOpen
  \bibfield  {author} {\bibinfo {author} {\bibfnamefont {K.}~\bibnamefont {Jedamzik}},\ }\href {https://doi.org/10.1103/PhysRevD.55.R5871} {\bibfield  {journal} {\bibinfo  {journal} {Phys. Rev. D}\ }\textbf {\bibinfo {volume} {55}},\ \bibinfo {pages} {5871} (\bibinfo {year} {1997})},\ \Eprint {https://arxiv.org/abs/astro-ph/9605152} {arXiv:astro-ph/9605152} \BibitemShut {NoStop}%
\bibitem [{\citenamefont {Jedamzik}\ and\ \citenamefont {Niemeyer}(1999)}]{Jedamzik:1999am}%
  \BibitemOpen
  \bibfield  {author} {\bibinfo {author} {\bibfnamefont {K.}~\bibnamefont {Jedamzik}}\ and\ \bibinfo {author} {\bibfnamefont {J.~C.}\ \bibnamefont {Niemeyer}},\ }\href {https://doi.org/10.1103/PhysRevD.59.124014} {\bibfield  {journal} {\bibinfo  {journal} {Phys. Rev. D}\ }\textbf {\bibinfo {volume} {59}},\ \bibinfo {pages} {124014} (\bibinfo {year} {1999})},\ \Eprint {https://arxiv.org/abs/astro-ph/9901293} {arXiv:astro-ph/9901293} \BibitemShut {NoStop}%
\bibitem [{\citenamefont {Byrnes}\ \emph {et~al.}(2018)\citenamefont {Byrnes}, \citenamefont {Hindmarsh}, \citenamefont {Young},\ and\ \citenamefont {Hawkins}}]{Byrnes:2018clq}%
  \BibitemOpen
  \bibfield  {author} {\bibinfo {author} {\bibfnamefont {C.~T.}\ \bibnamefont {Byrnes}}, \bibinfo {author} {\bibfnamefont {M.}~\bibnamefont {Hindmarsh}}, \bibinfo {author} {\bibfnamefont {S.}~\bibnamefont {Young}},\ and\ \bibinfo {author} {\bibfnamefont {M.~R.~S.}\ \bibnamefont {Hawkins}},\ }\href {https://doi.org/10.1088/1475-7516/2018/08/041} {\bibfield  {journal} {\bibinfo  {journal} {JCAP}\ }\textbf {\bibinfo {volume} {08}},\ \bibinfo {pages} {041}},\ \Eprint {https://arxiv.org/abs/1801.06138} {arXiv:1801.06138 [astro-ph.CO]} \BibitemShut {NoStop}%
\bibitem [{\citenamefont {Vovchenko}\ \emph {et~al.}(2021)\citenamefont {Vovchenko}, \citenamefont {Brandt}, \citenamefont {Cuteri}, \citenamefont {Endr{\H{o}}di}, \citenamefont {Hajkarim},\ and\ \citenamefont {Schaffner-Bielich}}]{Vovchenko:2020crk}%
  \BibitemOpen
  \bibfield  {author} {\bibinfo {author} {\bibfnamefont {V.}~\bibnamefont {Vovchenko}}, \bibinfo {author} {\bibfnamefont {B.~B.}\ \bibnamefont {Brandt}}, \bibinfo {author} {\bibfnamefont {F.}~\bibnamefont {Cuteri}}, \bibinfo {author} {\bibfnamefont {G.}~\bibnamefont {Endr{\H{o}}di}}, \bibinfo {author} {\bibfnamefont {F.}~\bibnamefont {Hajkarim}},\ and\ \bibinfo {author} {\bibfnamefont {J.}~\bibnamefont {Schaffner-Bielich}},\ }\href {https://doi.org/10.1103/PhysRevLett.126.012701} {\bibfield  {journal} {\bibinfo  {journal} {Phys. Rev. Lett.}\ }\textbf {\bibinfo {volume} {126}},\ \bibinfo {pages} {012701} (\bibinfo {year} {2021})},\ \Eprint {https://arxiv.org/abs/2009.02309} {arXiv:2009.02309 [hep-ph]} \BibitemShut {NoStop}%
\bibitem [{\citenamefont {B{\"o}deker}\ \emph {et~al.}(2021)\citenamefont {B{\"o}deker}, \citenamefont {K{\"u}hnel}, \citenamefont {Oldengott},\ and\ \citenamefont {Schwarz}}]{Bodeker:2020stj}%
  \BibitemOpen
  \bibfield  {author} {\bibinfo {author} {\bibfnamefont {D.}~\bibnamefont {B{\"o}deker}}, \bibinfo {author} {\bibfnamefont {F.}~\bibnamefont {K{\"u}hnel}}, \bibinfo {author} {\bibfnamefont {I.~M.}\ \bibnamefont {Oldengott}},\ and\ \bibinfo {author} {\bibfnamefont {D.~J.}\ \bibnamefont {Schwarz}},\ }\href {https://doi.org/10.1103/PhysRevD.103.063506} {\bibfield  {journal} {\bibinfo  {journal} {Phys. Rev. D}\ }\textbf {\bibinfo {volume} {103}},\ \bibinfo {pages} {063506} (\bibinfo {year} {2021})},\ \Eprint {https://arxiv.org/abs/2011.07283} {arXiv:2011.07283 [astro-ph.CO]} \BibitemShut {NoStop}%
\bibitem [{\citenamefont {Sinha}(2022)}]{Sinha:2022jfr}%
  \BibitemOpen
  \bibfield  {author} {\bibinfo {author} {\bibfnamefont {B.}~\bibnamefont {Sinha}},\ }\href {https://doi.org/10.1134/S1063779622020769} {\bibfield  {journal} {\bibinfo  {journal} {Phys. Part. Nucl.}\ }\textbf {\bibinfo {volume} {53}},\ \bibinfo {pages} {159} (\bibinfo {year} {2022})}\BibitemShut {NoStop}%
\bibitem [{\citenamefont {Di~Clemente}\ \emph {et~al.}(2025)\citenamefont {Di~Clemente}, \citenamefont {Casolino}, \citenamefont {Drago}, \citenamefont {Lattanzi},\ and\ \citenamefont {Ratti}}]{DiClemente:2024lzi}%
  \BibitemOpen
  \bibfield  {author} {\bibinfo {author} {\bibfnamefont {F.}~\bibnamefont {Di~Clemente}}, \bibinfo {author} {\bibfnamefont {M.}~\bibnamefont {Casolino}}, \bibinfo {author} {\bibfnamefont {A.}~\bibnamefont {Drago}}, \bibinfo {author} {\bibfnamefont {M.}~\bibnamefont {Lattanzi}},\ and\ \bibinfo {author} {\bibfnamefont {C.}~\bibnamefont {Ratti}},\ }\href {https://doi.org/10.1093/mnras/staf087} {\bibfield  {journal} {\bibinfo  {journal} {Mon. Not. Roy. Astron. Soc.}\ }\textbf {\bibinfo {volume} {537}},\ \bibinfo {pages} {1056} (\bibinfo {year} {2025})},\ \Eprint {https://arxiv.org/abs/2404.12094} {arXiv:2404.12094 [hep-ph]} \BibitemShut {NoStop}%
\bibitem [{\citenamefont {Gao}\ and\ \citenamefont {Oldengott}(2022{\natexlab{a}})}]{Gao:2021nwz}%
  \BibitemOpen
  \bibfield  {author} {\bibinfo {author} {\bibfnamefont {F.}~\bibnamefont {Gao}}\ and\ \bibinfo {author} {\bibfnamefont {I.~M.}\ \bibnamefont {Oldengott}},\ }\href {https://doi.org/10.1103/PhysRevLett.128.131301} {\bibfield  {journal} {\bibinfo  {journal} {Phys. Rev. Lett.}\ }\textbf {\bibinfo {volume} {128}},\ \bibinfo {pages} {131301} (\bibinfo {year} {2022}{\natexlab{a}})},\ \Eprint {https://arxiv.org/abs/2106.11991} {arXiv:2106.11991 [hep-ph]} \BibitemShut {NoStop}%
\bibitem [{\citenamefont {Ferreira}\ \emph {et~al.}(2025)\citenamefont {Ferreira}, \citenamefont {Fraga}, \citenamefont {Hippert},\ and\ \citenamefont {Schaffner-Bielich}}]{Ferreira:2025zeu}%
  \BibitemOpen
  \bibfield  {author} {\bibinfo {author} {\bibfnamefont {O.}~\bibnamefont {Ferreira}}, \bibinfo {author} {\bibfnamefont {E.~S.}\ \bibnamefont {Fraga}}, \bibinfo {author} {\bibfnamefont {M.}~\bibnamefont {Hippert}},\ and\ \bibinfo {author} {\bibfnamefont {J.}~\bibnamefont {Schaffner-Bielich}},\ }\href@noop {} {\bibinfo {title} {{Chiral symmetry breaking and pion condensation in the early universe}}} (\bibinfo {year} {2025}),\ \bibinfo {note} {arXiv:2507.06518},\ \Eprint {https://arxiv.org/abs/"2507.06518"} {arXiv:"2507.06518" [hep-ph]} \BibitemShut {NoStop}%
\bibitem [{\citenamefont {Gao}\ and\ \citenamefont {Oldengott}(2022{\natexlab{b}})}]{gao2022cosmology}%
  \BibitemOpen
  \bibfield  {author} {\bibinfo {author} {\bibfnamefont {F.}~\bibnamefont {Gao}}\ and\ \bibinfo {author} {\bibfnamefont {I.~M.}\ \bibnamefont {Oldengott}},\ }\href@noop {} {\bibfield  {journal} {\bibinfo  {journal} {Physical Review Letters}\ }\textbf {\bibinfo {volume} {128}},\ \bibinfo {pages} {131301} (\bibinfo {year} {2022}{\natexlab{b}})}\BibitemShut {NoStop}%
\bibitem [{\citenamefont {Wygas}\ \emph {et~al.}(2018{\natexlab{a}})\citenamefont {Wygas}, \citenamefont {Oldengott}, \citenamefont {B{\"o}deker},\ and\ \citenamefont {Schwarz}}]{wygas2018cosmic}%
  \BibitemOpen
  \bibfield  {author} {\bibinfo {author} {\bibfnamefont {M.~M.}\ \bibnamefont {Wygas}}, \bibinfo {author} {\bibfnamefont {I.~M.}\ \bibnamefont {Oldengott}}, \bibinfo {author} {\bibfnamefont {D.}~\bibnamefont {B{\"o}deker}},\ and\ \bibinfo {author} {\bibfnamefont {D.~J.}\ \bibnamefont {Schwarz}},\ }\href@noop {} {\bibfield  {journal} {\bibinfo  {journal} {Physical Review Letters}\ }\textbf {\bibinfo {volume} {121}},\ \bibinfo {pages} {201302} (\bibinfo {year} {2018}{\natexlab{a}})}\BibitemShut {NoStop}%
\bibitem [{\citenamefont {Sakharov}(1967)}]{Sakharov:1967dj}%
  \BibitemOpen
  \bibfield  {author} {\bibinfo {author} {\bibfnamefont {A.~D.}\ \bibnamefont {Sakharov}},\ }\href {https://doi.org/10.1070/PU1991v034n05ABEH002497} {\bibfield  {journal} {\bibinfo  {journal} {Pisma Zh. Eksp. Teor. Fiz.}\ }\textbf {\bibinfo {volume} {5}},\ \bibinfo {pages} {32} (\bibinfo {year} {1967})}\BibitemShut {NoStop}%
\bibitem [{\citenamefont {Barr}\ \emph {et~al.}(1979)\citenamefont {Barr}, \citenamefont {Segre},\ and\ \citenamefont {Weldon}}]{Barr:1979ye}%
  \BibitemOpen
  \bibfield  {author} {\bibinfo {author} {\bibfnamefont {S.~M.}\ \bibnamefont {Barr}}, \bibinfo {author} {\bibfnamefont {G.}~\bibnamefont {Segre}},\ and\ \bibinfo {author} {\bibfnamefont {H.~A.}\ \bibnamefont {Weldon}},\ }\href {https://doi.org/10.1103/PhysRevD.20.2494} {\bibfield  {journal} {\bibinfo  {journal} {Phys. Rev. D}\ }\textbf {\bibinfo {volume} {20}},\ \bibinfo {pages} {2494} (\bibinfo {year} {1979})}\BibitemShut {NoStop}%
\bibitem [{\citenamefont {Farrar}\ and\ \citenamefont {Shaposhnikov}(1994)}]{Farrar_1994}%
  \BibitemOpen
  \bibfield  {author} {\bibinfo {author} {\bibfnamefont {G.~R.}\ \bibnamefont {Farrar}}\ and\ \bibinfo {author} {\bibfnamefont {M.~E.}\ \bibnamefont {Shaposhnikov}},\ }\href {https://doi.org/10.1103/physrevd.50.774} {\bibfield  {journal} {\bibinfo  {journal} {Physical Review D}\ }\textbf {\bibinfo {volume} {50}},\ \bibinfo {pages} {774–818} (\bibinfo {year} {1994})}\BibitemShut {NoStop}%
\bibitem [{\citenamefont {Harvey}\ and\ \citenamefont {Turner}(1990)}]{PhysRevD.42.3344}%
  \BibitemOpen
  \bibfield  {author} {\bibinfo {author} {\bibfnamefont {J.~A.}\ \bibnamefont {Harvey}}\ and\ \bibinfo {author} {\bibfnamefont {M.~S.}\ \bibnamefont {Turner}},\ }\href {https://doi.org/10.1103/PhysRevD.42.3344} {\bibfield  {journal} {\bibinfo  {journal} {Phys. Rev. D}\ }\textbf {\bibinfo {volume} {42}},\ \bibinfo {pages} {3344} (\bibinfo {year} {1990})}\BibitemShut {NoStop}%
\bibitem [{\citenamefont {Eijima}\ and\ \citenamefont {Shaposhnikov}(2017)}]{Eijima:2017anv}%
  \BibitemOpen
  \bibfield  {author} {\bibinfo {author} {\bibfnamefont {S.}~\bibnamefont {Eijima}}\ and\ \bibinfo {author} {\bibfnamefont {M.}~\bibnamefont {Shaposhnikov}},\ }\href {https://doi.org/10.1016/j.physletb.2017.05.068} {\bibfield  {journal} {\bibinfo  {journal} {Phys. Lett. B}\ }\textbf {\bibinfo {volume} {771}},\ \bibinfo {pages} {288} (\bibinfo {year} {2017})},\ \Eprint {https://arxiv.org/abs/1703.06085} {arXiv:1703.06085 [hep-ph]} \BibitemShut {NoStop}%
\bibitem [{\citenamefont {Caprini}\ \emph {et~al.}(2005)\citenamefont {Caprini}, \citenamefont {Biller},\ and\ \citenamefont {Ferreira}}]{Caprini:2003gz}%
  \BibitemOpen
  \bibfield  {author} {\bibinfo {author} {\bibfnamefont {C.}~\bibnamefont {Caprini}}, \bibinfo {author} {\bibfnamefont {S.}~\bibnamefont {Biller}},\ and\ \bibinfo {author} {\bibfnamefont {P.~G.}\ \bibnamefont {Ferreira}},\ }\href {https://doi.org/10.1088/1475-7516/2005/02/006} {\bibfield  {journal} {\bibinfo  {journal} {JCAP}\ }\textbf {\bibinfo {volume} {02}},\ \bibinfo {pages} {006}},\ \Eprint {https://arxiv.org/abs/hep-ph/0310066} {arXiv:hep-ph/0310066} \BibitemShut {NoStop}%
\bibitem [{\citenamefont {Oldengott}\ and\ \citenamefont {Schwarz}(2017{\natexlab{a}})}]{Oldengott:2017tzj}%
  \BibitemOpen
  \bibfield  {author} {\bibinfo {author} {\bibfnamefont {I.~M.}\ \bibnamefont {Oldengott}}\ and\ \bibinfo {author} {\bibfnamefont {D.~J.}\ \bibnamefont {Schwarz}},\ }\href {https://doi.org/10.1209/0295-5075/119/29001} {\bibfield  {journal} {\bibinfo  {journal} {EPL}\ }\textbf {\bibinfo {volume} {119}},\ \bibinfo {pages} {29001} (\bibinfo {year} {2017}{\natexlab{a}})},\ \Eprint {https://arxiv.org/abs/1706.01705} {arXiv:1706.01705 [astro-ph.CO]} \BibitemShut {NoStop}%
\bibitem [{\citenamefont {Mangano}\ \emph {et~al.}(2012{\natexlab{a}})\citenamefont {Mangano}, \citenamefont {Miele}, \citenamefont {Pastor}, \citenamefont {Pisanti},\ and\ \citenamefont {Sarikas}}]{Mangano:2011ip}%
  \BibitemOpen
  \bibfield  {author} {\bibinfo {author} {\bibfnamefont {G.}~\bibnamefont {Mangano}}, \bibinfo {author} {\bibfnamefont {G.}~\bibnamefont {Miele}}, \bibinfo {author} {\bibfnamefont {S.}~\bibnamefont {Pastor}}, \bibinfo {author} {\bibfnamefont {O.}~\bibnamefont {Pisanti}},\ and\ \bibinfo {author} {\bibfnamefont {S.}~\bibnamefont {Sarikas}},\ }\href {https://doi.org/10.1016/j.physletb.2012.01.015} {\bibfield  {journal} {\bibinfo  {journal} {Phys. Lett. B}\ }\textbf {\bibinfo {volume} {708}},\ \bibinfo {pages} {1} (\bibinfo {year} {2012}{\natexlab{a}})},\ \Eprint {https://arxiv.org/abs/1110.4335} {arXiv:1110.4335 [hep-ph]} \BibitemShut {NoStop}%
\bibitem [{\citenamefont {Bzdak}\ \emph {et~al.}(2020)\citenamefont {Bzdak}, \citenamefont {Esumi}, \citenamefont {Koch}, \citenamefont {Liao}, \citenamefont {Stephanov},\ and\ \citenamefont {Xu}}]{Bzdak:2019pkr}%
  \BibitemOpen
  \bibfield  {author} {\bibinfo {author} {\bibfnamefont {A.}~\bibnamefont {Bzdak}}, \bibinfo {author} {\bibfnamefont {S.}~\bibnamefont {Esumi}}, \bibinfo {author} {\bibfnamefont {V.}~\bibnamefont {Koch}}, \bibinfo {author} {\bibfnamefont {J.}~\bibnamefont {Liao}}, \bibinfo {author} {\bibfnamefont {M.}~\bibnamefont {Stephanov}},\ and\ \bibinfo {author} {\bibfnamefont {N.}~\bibnamefont {Xu}},\ }\href {https://doi.org/10.1016/j.physrep.2020.01.005} {\bibfield  {journal} {\bibinfo  {journal} {Phys. Rept.}\ }\textbf {\bibinfo {volume} {853}},\ \bibinfo {pages} {1} (\bibinfo {year} {2020})},\ \Eprint {https://arxiv.org/abs/1906.00936} {arXiv:1906.00936 [nucl-th]} \BibitemShut {NoStop}%
\bibitem [{\citenamefont {Du}\ \emph {et~al.}(2024)\citenamefont {Du}, \citenamefont {Sorensen},\ and\ \citenamefont {Stephanov}}]{Du:2024wjm}%
  \BibitemOpen
  \bibfield  {author} {\bibinfo {author} {\bibfnamefont {L.}~\bibnamefont {Du}}, \bibinfo {author} {\bibfnamefont {A.}~\bibnamefont {Sorensen}},\ and\ \bibinfo {author} {\bibfnamefont {M.}~\bibnamefont {Stephanov}},\ }\href {https://doi.org/10.1142/9789811294679_0007} {\bibfield  {journal} {\bibinfo  {journal} {Int. J. Mod. Phys. E}\ }\textbf {\bibinfo {volume} {33}},\ \bibinfo {pages} {2430008} (\bibinfo {year} {2024})},\ \Eprint {https://arxiv.org/abs/2402.10183} {arXiv:2402.10183 [nucl-th]} \BibitemShut {NoStop}%
\bibitem [{\citenamefont {Fu}\ \emph {et~al.}(2020)\citenamefont {Fu}, \citenamefont {Pawlowski},\ and\ \citenamefont {Rennecke}}]{Fu:2019hdw}%
  \BibitemOpen
  \bibfield  {author} {\bibinfo {author} {\bibfnamefont {W.-j.}\ \bibnamefont {Fu}}, \bibinfo {author} {\bibfnamefont {J.~M.}\ \bibnamefont {Pawlowski}},\ and\ \bibinfo {author} {\bibfnamefont {F.}~\bibnamefont {Rennecke}},\ }\href {https://doi.org/10.1103/PhysRevD.101.054032} {\bibfield  {journal} {\bibinfo  {journal} {Phys. Rev. D}\ }\textbf {\bibinfo {volume} {101}},\ \bibinfo {pages} {054032} (\bibinfo {year} {2020})},\ \Eprint {https://arxiv.org/abs/1909.02991} {arXiv:1909.02991 [hep-ph]} \BibitemShut {NoStop}%
\bibitem [{\citenamefont {Gunkel}\ and\ \citenamefont {Fischer}(2021)}]{Gunkel:2021oya}%
  \BibitemOpen
  \bibfield  {author} {\bibinfo {author} {\bibfnamefont {P.~J.}\ \bibnamefont {Gunkel}}\ and\ \bibinfo {author} {\bibfnamefont {C.~S.}\ \bibnamefont {Fischer}},\ }\href {https://doi.org/10.1103/PhysRevD.104.054022} {\bibfield  {journal} {\bibinfo  {journal} {Phys. Rev. D}\ }\textbf {\bibinfo {volume} {104}},\ \bibinfo {pages} {054022} (\bibinfo {year} {2021})},\ \Eprint {https://arxiv.org/abs/2106.08356} {arXiv:2106.08356 [hep-ph]} \BibitemShut {NoStop}%
\bibitem [{\citenamefont {Gao}\ and\ \citenamefont {Pawlowski}(2021)}]{Gao:2020fbl}%
  \BibitemOpen
  \bibfield  {author} {\bibinfo {author} {\bibfnamefont {F.}~\bibnamefont {Gao}}\ and\ \bibinfo {author} {\bibfnamefont {J.~M.}\ \bibnamefont {Pawlowski}},\ }\href {https://doi.org/10.1016/j.physletb.2021.136584} {\bibfield  {journal} {\bibinfo  {journal} {Phys. Lett. B}\ }\textbf {\bibinfo {volume} {820}},\ \bibinfo {pages} {136584} (\bibinfo {year} {2021})},\ \Eprint {https://arxiv.org/abs/2010.13705} {arXiv:2010.13705 [hep-ph]} \BibitemShut {NoStop}%
\bibitem [{\citenamefont {Hippert}\ \emph {et~al.}(2024)\citenamefont {Hippert}, \citenamefont {Grefa}, \citenamefont {Manning}, \citenamefont {Noronha}, \citenamefont {Noronha-Hostler}, \citenamefont {Portillo~Vazquez}, \citenamefont {Ratti}, \citenamefont {Rougemont},\ and\ \citenamefont {Trujillo}}]{Hippert:2023bel}%
  \BibitemOpen
  \bibfield  {author} {\bibinfo {author} {\bibfnamefont {M.}~\bibnamefont {Hippert}}, \bibinfo {author} {\bibfnamefont {J.}~\bibnamefont {Grefa}}, \bibinfo {author} {\bibfnamefont {T.~A.}\ \bibnamefont {Manning}}, \bibinfo {author} {\bibfnamefont {J.}~\bibnamefont {Noronha}}, \bibinfo {author} {\bibfnamefont {J.}~\bibnamefont {Noronha-Hostler}}, \bibinfo {author} {\bibfnamefont {I.}~\bibnamefont {Portillo~Vazquez}}, \bibinfo {author} {\bibfnamefont {C.}~\bibnamefont {Ratti}}, \bibinfo {author} {\bibfnamefont {R.}~\bibnamefont {Rougemont}},\ and\ \bibinfo {author} {\bibfnamefont {M.}~\bibnamefont {Trujillo}},\ }\href {https://doi.org/10.1103/PhysRevD.110.094006} {\bibfield  {journal} {\bibinfo  {journal} {Phys. Rev. D}\ }\textbf {\bibinfo {volume} {110}},\ \bibinfo {pages} {094006} (\bibinfo {year} {2024})},\ \Eprint {https://arxiv.org/abs/2309.00579} {arXiv:2309.00579 [nucl-th]} \BibitemShut {NoStop}%
\bibitem [{\citenamefont {Basar}(2024)}]{Basar:2023nkp}%
  \BibitemOpen
  \bibfield  {author} {\bibinfo {author} {\bibfnamefont {G.}~\bibnamefont {Basar}},\ }\href {https://doi.org/10.1103/PhysRevC.110.015203} {\bibfield  {journal} {\bibinfo  {journal} {Phys. Rev. C}\ }\textbf {\bibinfo {volume} {110}},\ \bibinfo {pages} {015203} (\bibinfo {year} {2024})},\ \Eprint {https://arxiv.org/abs/2312.06952} {arXiv:2312.06952 [hep-th]} \BibitemShut {NoStop}%
\bibitem [{\citenamefont {Clarke}\ \emph {et~al.}(2024)\citenamefont {Clarke}, \citenamefont {Dimopoulos}, \citenamefont {Di~Renzo}, \citenamefont {Goswami}, \citenamefont {Schmidt}, \citenamefont {Singh},\ and\ \citenamefont {Zambello}}]{Clarke:2024ugt}%
  \BibitemOpen
  \bibfield  {author} {\bibinfo {author} {\bibfnamefont {D.~A.}\ \bibnamefont {Clarke}}, \bibinfo {author} {\bibfnamefont {P.}~\bibnamefont {Dimopoulos}}, \bibinfo {author} {\bibfnamefont {F.}~\bibnamefont {Di~Renzo}}, \bibinfo {author} {\bibfnamefont {J.}~\bibnamefont {Goswami}}, \bibinfo {author} {\bibfnamefont {C.}~\bibnamefont {Schmidt}}, \bibinfo {author} {\bibfnamefont {S.}~\bibnamefont {Singh}},\ and\ \bibinfo {author} {\bibfnamefont {K.}~\bibnamefont {Zambello}},\ }\href@noop {} {\bibinfo {title} {{Searching for the QCD critical endpoint using multi-point Pad{\'e} approximations}}} (\bibinfo {year} {2024}),\ \Eprint {https://arxiv.org/abs/2405.10196} {arXiv:2405.10196 [hep-lat]} \BibitemShut {NoStop}%
\bibitem [{\citenamefont {Shah}\ \emph {et~al.}(2024)\citenamefont {Shah}, \citenamefont {Hippert}, \citenamefont {Noronha}, \citenamefont {Ratti},\ and\ \citenamefont {Vovchenko}}]{Shah:2024img}%
  \BibitemOpen
  \bibfield  {author} {\bibinfo {author} {\bibfnamefont {H.}~\bibnamefont {Shah}}, \bibinfo {author} {\bibfnamefont {M.}~\bibnamefont {Hippert}}, \bibinfo {author} {\bibfnamefont {J.}~\bibnamefont {Noronha}}, \bibinfo {author} {\bibfnamefont {C.}~\bibnamefont {Ratti}},\ and\ \bibinfo {author} {\bibfnamefont {V.}~\bibnamefont {Vovchenko}},\ }\href@noop {} {\bibinfo {title} {{Locating the QCD critical point from first principles through contours of constant entropy density}}} (\bibinfo {year} {2024}),\ \Eprint {https://arxiv.org/abs/2410.16206} {arXiv:2410.16206 [hep-ph]} \BibitemShut {NoStop}%
\bibitem [{\citenamefont {Borsanyi}\ \emph {et~al.}(2020)\citenamefont {Borsanyi}, \citenamefont {Fodor}, \citenamefont {Guenther}, \citenamefont {Kara}, \citenamefont {Katz}, \citenamefont {Parotto}, \citenamefont {Pasztor}, \citenamefont {Ratti},\ and\ \citenamefont {Szabo}}]{Borsanyi:2020fev}%
  \BibitemOpen
  \bibfield  {author} {\bibinfo {author} {\bibfnamefont {S.}~\bibnamefont {Borsanyi}}, \bibinfo {author} {\bibfnamefont {Z.}~\bibnamefont {Fodor}}, \bibinfo {author} {\bibfnamefont {J.~N.}\ \bibnamefont {Guenther}}, \bibinfo {author} {\bibfnamefont {R.}~\bibnamefont {Kara}}, \bibinfo {author} {\bibfnamefont {S.~D.}\ \bibnamefont {Katz}}, \bibinfo {author} {\bibfnamefont {P.}~\bibnamefont {Parotto}}, \bibinfo {author} {\bibfnamefont {A.}~\bibnamefont {Pasztor}}, \bibinfo {author} {\bibfnamefont {C.}~\bibnamefont {Ratti}},\ and\ \bibinfo {author} {\bibfnamefont {K.~K.}\ \bibnamefont {Szabo}},\ }\href {https://doi.org/10.1103/PhysRevLett.125.052001} {\bibfield  {journal} {\bibinfo  {journal} {Phys. Rev. Lett.}\ }\textbf {\bibinfo {volume} {125}},\ \bibinfo {pages} {052001} (\bibinfo {year} {2020})},\ \Eprint {https://arxiv.org/abs/2002.02821} {arXiv:2002.02821 [hep-lat]} \BibitemShut {NoStop}%
\bibitem [{\citenamefont {{Planck Collaboration}}\ \emph {et~al.}(2020)\citenamefont {{Planck Collaboration}}, \citenamefont {{Aghanim, N.}}, \citenamefont {{Akrami, Y.}}, \citenamefont {{Ashdown, M.}}, \citenamefont {{Aumont, J.}}, \citenamefont {{Baccigalupi, C.}}, \citenamefont {{Ballardini, M.}}, \citenamefont {{Banday, A. J.}}, \citenamefont {{Barreiro, R. B.}}, \citenamefont {{Bartolo, N.}}, \citenamefont {{Basak, S.}}, \citenamefont {{Battye, R.}}, \citenamefont {{Benabed, K.}}, \citenamefont {{Bernard, J.-P.}}, \citenamefont {{Bersanelli, M.}}, \citenamefont {{Bielewicz, P.}}, \citenamefont {{Bock, J. J.}}, \citenamefont {{Bond, J. R.}}, \citenamefont {{Borrill, J.}}, \citenamefont {{Bouchet, F. R.}}, \citenamefont {{Boulanger, F.}}, \citenamefont {{Bucher, M.}}, \citenamefont {{Burigana, C.}}, \citenamefont {{Butler, R. C.}}, \citenamefont {{Calabrese, E.}}, \citenamefont {{Cardoso, J.-F.}}, \citenamefont {{Carron, J.}}, \citenamefont {{Challinor, A.}}, \citenamefont {{Chiang, H. C.}}, \citenamefont
  {{Chluba, J.}}, \citenamefont {{Colombo, L. P. L.}}, \citenamefont {{Combet, C.}}, \citenamefont {{Contreras, D.}}, \citenamefont {{Crill, B. P.}}, \citenamefont {{Cuttaia, F.}}, \citenamefont {{de Bernardis, P.}}, \citenamefont {{de Zotti, G.}}, \citenamefont {{Delabrouille, J.}}, \citenamefont {{Delouis, J.-M.}}, \citenamefont {{Di Valentino, E.}}, \citenamefont {{Diego, J. M.}}, \citenamefont {{Doré, O.}}, \citenamefont {{Douspis, M.}}, \citenamefont {{Ducout, A.}}, \citenamefont {{Dupac, X.}}, \citenamefont {{Dusini, S.}}, \citenamefont {{Efstathiou, G.}}, \citenamefont {{Elsner, F.}}, \citenamefont {{Enßlin, T. A.}}, \citenamefont {{Eriksen, H. K.}}, \citenamefont {{Fantaye, Y.}}, \citenamefont {{Farhang, M.}}, \citenamefont {{Fergusson, J.}}, \citenamefont {{Fernandez-Cobos, R.}}, \citenamefont {{Finelli, F.}}, \citenamefont {{Forastieri, F.}}, \citenamefont {{Frailis, M.}}, \citenamefont {{Fraisse, A. A.}}, \citenamefont {{Franceschi, E.}}, \citenamefont {{Frolov, A.}}, \citenamefont {{Galeotta,
  S.}}, \citenamefont {{Galli, S.}}, \citenamefont {{Ganga, K.}}, \citenamefont {{Génova-Santos, R. T.}}, \citenamefont {{Gerbino, M.}}, \citenamefont {{Ghosh, T.}}, \citenamefont {{González-Nuevo, J.}}, \citenamefont {{Górski, K. M.}}, \citenamefont {{Gratton, S.}}, \citenamefont {{Gruppuso, A.}}, \citenamefont {{Gudmundsson, J. E.}}, \citenamefont {{Hamann, J.}}, \citenamefont {{Handley, W.}}, \citenamefont {{Hansen, F. K.}}, \citenamefont {{Herranz, D.}}, \citenamefont {{Hildebrandt, S. R.}}, \citenamefont {{Hivon, E.}}, \citenamefont {{Huang, Z.}}, \citenamefont {{Jaffe, A. H.}}, \citenamefont {{Jones, W. C.}}, \citenamefont {{Karakci, A.}}, \citenamefont {{Keihänen, E.}}, \citenamefont {{Keskitalo, R.}}, \citenamefont {{Kiiveri, K.}}, \citenamefont {{Kim, J.}}, \citenamefont {{Kisner, T. S.}}, \citenamefont {{Knox, L.}}, \citenamefont {{Krachmalnicoff, N.}}, \citenamefont {{Kunz, M.}}, \citenamefont {{Kurki-Suonio, H.}}, \citenamefont {{Lagache, G.}}, \citenamefont {{Lamarre, J.-M.}}, \citenamefont
  {{Lasenby, A.}}, \citenamefont {{Lattanzi, M.}}, \citenamefont {{Lawrence, C. R.}}, \citenamefont {{Le Jeune, M.}}, \citenamefont {{Lemos, P.}}, \citenamefont {{Lesgourgues, J.}}, \citenamefont {{Levrier, F.}}, \citenamefont {{Lewis, A.}}, \citenamefont {{Liguori, M.}}, \citenamefont {{Lilje, P. B.}}, \citenamefont {{Lilley, M.}}, \citenamefont {{Lindholm, V.}}, \citenamefont {{López-Caniego, M.}}, \citenamefont {{Lubin, P. M.}}, \citenamefont {{Ma, Y.-Z.}}, \citenamefont {{Macías-Pérez, J. F.}}, \citenamefont {{Maggio, G.}}, \citenamefont {{Maino, D.}}, \citenamefont {{Mandolesi, N.}}, \citenamefont {{Mangilli, A.}}, \citenamefont {{Marcos-Caballero, A.}}, \citenamefont {{Maris, M.}}, \citenamefont {{Martin, P. G.}}, \citenamefont {{Martinelli, M.}}, \citenamefont {{Martínez-González, E.}}, \citenamefont {{Matarrese, S.}}, \citenamefont {{Mauri, N.}}, \citenamefont {{McEwen, J. D.}}, \citenamefont {{Meinhold, P. R.}}, \citenamefont {{Melchiorri, A.}}, \citenamefont {{Mennella, A.}}, \citenamefont
  {{Migliaccio, M.}}, \citenamefont {{Millea, M.}}, \citenamefont {{Mitra, S.}}, \citenamefont {{Miville-Deschênes, M.-A.}}, \citenamefont {{Molinari, D.}}, \citenamefont {{Montier, L.}}, \citenamefont {{Morgante, G.}}, \citenamefont {{Moss, A.}}, \citenamefont {{Natoli, P.}}, \citenamefont {{Nørgaard-Nielsen, H. U.}}, \citenamefont {{Pagano, L.}}, \citenamefont {{Paoletti, D.}}, \citenamefont {{Partridge, B.}}, \citenamefont {{Patanchon, G.}}, \citenamefont {{Peiris, H. V.}}, \citenamefont {{Perrotta, F.}}, \citenamefont {{Pettorino, V.}}, \citenamefont {{Piacentini, F.}}, \citenamefont {{Polastri, L.}}, \citenamefont {{Polenta, G.}}, \citenamefont {{Puget, J.-L.}}, \citenamefont {{Rachen, J. P.}}, \citenamefont {{Reinecke, M.}}, \citenamefont {{Remazeilles, M.}}, \citenamefont {{Renzi, A.}}, \citenamefont {{Rocha, G.}}, \citenamefont {{Rosset, C.}}, \citenamefont {{Roudier, G.}}, \citenamefont {{Rubiño-Martín, J. A.}}, \citenamefont {{Ruiz-Granados, B.}}, \citenamefont {{Salvati, L.}}, \citenamefont
  {{Sandri, M.}}, \citenamefont {{Savelainen, M.}}, \citenamefont {{Scott, D.}}, \citenamefont {{Shellard, E. P. S.}}, \citenamefont {{Sirignano, C.}}, \citenamefont {{Sirri, G.}}, \citenamefont {{Spencer, L. D.}}, \citenamefont {{Sunyaev, R.}}, \citenamefont {{Suur-Uski, A.-S.}}, \citenamefont {{Tauber, J. A.}}, \citenamefont {{Tavagnacco, D.}}, \citenamefont {{Tenti, M.}}, \citenamefont {{Toffolatti, L.}}, \citenamefont {{Tomasi, M.}}, \citenamefont {{Trombetti, T.}}, \citenamefont {{Valenziano, L.}}, \citenamefont {{Valiviita, J.}}, \citenamefont {{Van Tent, B.}}, \citenamefont {{Vibert, L.}}, \citenamefont {{Vielva, P.}}, \citenamefont {{Villa, F.}}, \citenamefont {{Vittorio, N.}}, \citenamefont {{Wandelt, B. D.}}, \citenamefont {{Wehus, I. K.}}, \citenamefont {{White, M.}}, \citenamefont {{White, S. D. M.}}, \citenamefont {{Zacchei, A.}},\ and\ \citenamefont {{Zonca, A.}}}]{refId0}%
  \BibitemOpen
  \bibfield  {author} {\bibinfo {author} {\bibnamefont {{Planck Collaboration}}}, \bibinfo {author} {\bibnamefont {{Aghanim, N.}}}, \bibinfo {author} {\bibnamefont {{Akrami, Y.}}}, \bibinfo {author} {\bibnamefont {{Ashdown, M.}}}, \bibinfo {author} {\bibnamefont {{Aumont, J.}}}, \bibinfo {author} {\bibnamefont {{Baccigalupi, C.}}}, \bibinfo {author} {\bibnamefont {{Ballardini, M.}}}, \bibinfo {author} {\bibnamefont {{Banday, A. J.}}}, \bibinfo {author} {\bibnamefont {{Barreiro, R. B.}}}, \bibinfo {author} {\bibnamefont {{Bartolo, N.}}}, \bibinfo {author} {\bibnamefont {{Basak, S.}}}, \bibinfo {author} {\bibnamefont {{Battye, R.}}}, \bibinfo {author} {\bibnamefont {{Benabed, K.}}}, \bibinfo {author} {\bibnamefont {{Bernard, J.-P.}}}, \bibinfo {author} {\bibnamefont {{Bersanelli, M.}}}, \bibinfo {author} {\bibnamefont {{Bielewicz, P.}}}, \bibinfo {author} {\bibnamefont {{Bock, J. J.}}}, \bibinfo {author} {\bibnamefont {{Bond, J. R.}}}, \bibinfo {author} {\bibnamefont {{Borrill, J.}}}, \bibinfo {author}
  {\bibnamefont {{Bouchet, F. R.}}}, \bibinfo {author} {\bibnamefont {{Boulanger, F.}}}, \bibinfo {author} {\bibnamefont {{Bucher, M.}}}, \bibinfo {author} {\bibnamefont {{Burigana, C.}}}, \bibinfo {author} {\bibnamefont {{Butler, R. C.}}}, \bibinfo {author} {\bibnamefont {{Calabrese, E.}}}, \bibinfo {author} {\bibnamefont {{Cardoso, J.-F.}}}, \bibinfo {author} {\bibnamefont {{Carron, J.}}}, \bibinfo {author} {\bibnamefont {{Challinor, A.}}}, \bibinfo {author} {\bibnamefont {{Chiang, H. C.}}}, \bibinfo {author} {\bibnamefont {{Chluba, J.}}}, \bibinfo {author} {\bibnamefont {{Colombo, L. P. L.}}}, \bibinfo {author} {\bibnamefont {{Combet, C.}}}, \bibinfo {author} {\bibnamefont {{Contreras, D.}}}, \bibinfo {author} {\bibnamefont {{Crill, B. P.}}}, \bibinfo {author} {\bibnamefont {{Cuttaia, F.}}}, \bibinfo {author} {\bibnamefont {{de Bernardis, P.}}}, \bibinfo {author} {\bibnamefont {{de Zotti, G.}}}, \bibinfo {author} {\bibnamefont {{Delabrouille, J.}}}, \bibinfo {author} {\bibnamefont {{Delouis, J.-M.}}},
  \bibinfo {author} {\bibnamefont {{Di Valentino, E.}}}, \bibinfo {author} {\bibnamefont {{Diego, J. M.}}}, \bibinfo {author} {\bibnamefont {{Doré, O.}}}, \bibinfo {author} {\bibnamefont {{Douspis, M.}}}, \bibinfo {author} {\bibnamefont {{Ducout, A.}}}, \bibinfo {author} {\bibnamefont {{Dupac, X.}}}, \bibinfo {author} {\bibnamefont {{Dusini, S.}}}, \bibinfo {author} {\bibnamefont {{Efstathiou, G.}}}, \bibinfo {author} {\bibnamefont {{Elsner, F.}}}, \bibinfo {author} {\bibnamefont {{Enßlin, T. A.}}}, \bibinfo {author} {\bibnamefont {{Eriksen, H. K.}}}, \bibinfo {author} {\bibnamefont {{Fantaye, Y.}}}, \bibinfo {author} {\bibnamefont {{Farhang, M.}}}, \bibinfo {author} {\bibnamefont {{Fergusson, J.}}}, \bibinfo {author} {\bibnamefont {{Fernandez-Cobos, R.}}}, \bibinfo {author} {\bibnamefont {{Finelli, F.}}}, \bibinfo {author} {\bibnamefont {{Forastieri, F.}}}, \bibinfo {author} {\bibnamefont {{Frailis, M.}}}, \bibinfo {author} {\bibnamefont {{Fraisse, A. A.}}}, \bibinfo {author} {\bibnamefont {{Franceschi,
  E.}}}, \bibinfo {author} {\bibnamefont {{Frolov, A.}}}, \bibinfo {author} {\bibnamefont {{Galeotta, S.}}}, \bibinfo {author} {\bibnamefont {{Galli, S.}}}, \bibinfo {author} {\bibnamefont {{Ganga, K.}}}, \bibinfo {author} {\bibnamefont {{Génova-Santos, R. T.}}}, \bibinfo {author} {\bibnamefont {{Gerbino, M.}}}, \bibinfo {author} {\bibnamefont {{Ghosh, T.}}}, \bibinfo {author} {\bibnamefont {{González-Nuevo, J.}}}, \bibinfo {author} {\bibnamefont {{Górski, K. M.}}}, \bibinfo {author} {\bibnamefont {{Gratton, S.}}}, \bibinfo {author} {\bibnamefont {{Gruppuso, A.}}}, \bibinfo {author} {\bibnamefont {{Gudmundsson, J. E.}}}, \bibinfo {author} {\bibnamefont {{Hamann, J.}}}, \bibinfo {author} {\bibnamefont {{Handley, W.}}}, \bibinfo {author} {\bibnamefont {{Hansen, F. K.}}}, \bibinfo {author} {\bibnamefont {{Herranz, D.}}}, \bibinfo {author} {\bibnamefont {{Hildebrandt, S. R.}}}, \bibinfo {author} {\bibnamefont {{Hivon, E.}}}, \bibinfo {author} {\bibnamefont {{Huang, Z.}}}, \bibinfo {author} {\bibnamefont
  {{Jaffe, A. H.}}}, \bibinfo {author} {\bibnamefont {{Jones, W. C.}}}, \bibinfo {author} {\bibnamefont {{Karakci, A.}}}, \bibinfo {author} {\bibnamefont {{Keihänen, E.}}}, \bibinfo {author} {\bibnamefont {{Keskitalo, R.}}}, \bibinfo {author} {\bibnamefont {{Kiiveri, K.}}}, \bibinfo {author} {\bibnamefont {{Kim, J.}}}, \bibinfo {author} {\bibnamefont {{Kisner, T. S.}}}, \bibinfo {author} {\bibnamefont {{Knox, L.}}}, \bibinfo {author} {\bibnamefont {{Krachmalnicoff, N.}}}, \bibinfo {author} {\bibnamefont {{Kunz, M.}}}, \bibinfo {author} {\bibnamefont {{Kurki-Suonio, H.}}}, \bibinfo {author} {\bibnamefont {{Lagache, G.}}}, \bibinfo {author} {\bibnamefont {{Lamarre, J.-M.}}}, \bibinfo {author} {\bibnamefont {{Lasenby, A.}}}, \bibinfo {author} {\bibnamefont {{Lattanzi, M.}}}, \bibinfo {author} {\bibnamefont {{Lawrence, C. R.}}}, \bibinfo {author} {\bibnamefont {{Le Jeune, M.}}}, \bibinfo {author} {\bibnamefont {{Lemos, P.}}}, \bibinfo {author} {\bibnamefont {{Lesgourgues, J.}}}, \bibinfo {author} {\bibnamefont
  {{Levrier, F.}}}, \bibinfo {author} {\bibnamefont {{Lewis, A.}}}, \bibinfo {author} {\bibnamefont {{Liguori, M.}}}, \bibinfo {author} {\bibnamefont {{Lilje, P. B.}}}, \bibinfo {author} {\bibnamefont {{Lilley, M.}}}, \bibinfo {author} {\bibnamefont {{Lindholm, V.}}}, \bibinfo {author} {\bibnamefont {{López-Caniego, M.}}}, \bibinfo {author} {\bibnamefont {{Lubin, P. M.}}}, \bibinfo {author} {\bibnamefont {{Ma, Y.-Z.}}}, \bibinfo {author} {\bibnamefont {{Macías-Pérez, J. F.}}}, \bibinfo {author} {\bibnamefont {{Maggio, G.}}}, \bibinfo {author} {\bibnamefont {{Maino, D.}}}, \bibinfo {author} {\bibnamefont {{Mandolesi, N.}}}, \bibinfo {author} {\bibnamefont {{Mangilli, A.}}}, \bibinfo {author} {\bibnamefont {{Marcos-Caballero, A.}}}, \bibinfo {author} {\bibnamefont {{Maris, M.}}}, \bibinfo {author} {\bibnamefont {{Martin, P. G.}}}, \bibinfo {author} {\bibnamefont {{Martinelli, M.}}}, \bibinfo {author} {\bibnamefont {{Martínez-González, E.}}}, \bibinfo {author} {\bibnamefont {{Matarrese, S.}}}, \bibinfo
  {author} {\bibnamefont {{Mauri, N.}}}, \bibinfo {author} {\bibnamefont {{McEwen, J. D.}}}, \bibinfo {author} {\bibnamefont {{Meinhold, P. R.}}}, \bibinfo {author} {\bibnamefont {{Melchiorri, A.}}}, \bibinfo {author} {\bibnamefont {{Mennella, A.}}}, \bibinfo {author} {\bibnamefont {{Migliaccio, M.}}}, \bibinfo {author} {\bibnamefont {{Millea, M.}}}, \bibinfo {author} {\bibnamefont {{Mitra, S.}}}, \bibinfo {author} {\bibnamefont {{Miville-Deschênes, M.-A.}}}, \bibinfo {author} {\bibnamefont {{Molinari, D.}}}, \bibinfo {author} {\bibnamefont {{Montier, L.}}}, \bibinfo {author} {\bibnamefont {{Morgante, G.}}}, \bibinfo {author} {\bibnamefont {{Moss, A.}}}, \bibinfo {author} {\bibnamefont {{Natoli, P.}}}, \bibinfo {author} {\bibnamefont {{Nørgaard-Nielsen, H. U.}}}, \bibinfo {author} {\bibnamefont {{Pagano, L.}}}, \bibinfo {author} {\bibnamefont {{Paoletti, D.}}}, \bibinfo {author} {\bibnamefont {{Partridge, B.}}}, \bibinfo {author} {\bibnamefont {{Patanchon, G.}}}, \bibinfo {author} {\bibnamefont {{Peiris, H.
  V.}}}, \bibinfo {author} {\bibnamefont {{Perrotta, F.}}}, \bibinfo {author} {\bibnamefont {{Pettorino, V.}}}, \bibinfo {author} {\bibnamefont {{Piacentini, F.}}}, \bibinfo {author} {\bibnamefont {{Polastri, L.}}}, \bibinfo {author} {\bibnamefont {{Polenta, G.}}}, \bibinfo {author} {\bibnamefont {{Puget, J.-L.}}}, \bibinfo {author} {\bibnamefont {{Rachen, J. P.}}}, \bibinfo {author} {\bibnamefont {{Reinecke, M.}}}, \bibinfo {author} {\bibnamefont {{Remazeilles, M.}}}, \bibinfo {author} {\bibnamefont {{Renzi, A.}}}, \bibinfo {author} {\bibnamefont {{Rocha, G.}}}, \bibinfo {author} {\bibnamefont {{Rosset, C.}}}, \bibinfo {author} {\bibnamefont {{Roudier, G.}}}, \bibinfo {author} {\bibnamefont {{Rubiño-Martín, J. A.}}}, \bibinfo {author} {\bibnamefont {{Ruiz-Granados, B.}}}, \bibinfo {author} {\bibnamefont {{Salvati, L.}}}, \bibinfo {author} {\bibnamefont {{Sandri, M.}}}, \bibinfo {author} {\bibnamefont {{Savelainen, M.}}}, \bibinfo {author} {\bibnamefont {{Scott, D.}}}, \bibinfo {author} {\bibnamefont
  {{Shellard, E. P. S.}}}, \bibinfo {author} {\bibnamefont {{Sirignano, C.}}}, \bibinfo {author} {\bibnamefont {{Sirri, G.}}}, \bibinfo {author} {\bibnamefont {{Spencer, L. D.}}}, \bibinfo {author} {\bibnamefont {{Sunyaev, R.}}}, \bibinfo {author} {\bibnamefont {{Suur-Uski, A.-S.}}}, \bibinfo {author} {\bibnamefont {{Tauber, J. A.}}}, \bibinfo {author} {\bibnamefont {{Tavagnacco, D.}}}, \bibinfo {author} {\bibnamefont {{Tenti, M.}}}, \bibinfo {author} {\bibnamefont {{Toffolatti, L.}}}, \bibinfo {author} {\bibnamefont {{Tomasi, M.}}}, \bibinfo {author} {\bibnamefont {{Trombetti, T.}}}, \bibinfo {author} {\bibnamefont {{Valenziano, L.}}}, \bibinfo {author} {\bibnamefont {{Valiviita, J.}}}, \bibinfo {author} {\bibnamefont {{Van Tent, B.}}}, \bibinfo {author} {\bibnamefont {{Vibert, L.}}}, \bibinfo {author} {\bibnamefont {{Vielva, P.}}}, \bibinfo {author} {\bibnamefont {{Villa, F.}}}, \bibinfo {author} {\bibnamefont {{Vittorio, N.}}}, \bibinfo {author} {\bibnamefont {{Wandelt, B. D.}}}, \bibinfo {author}
  {\bibnamefont {{Wehus, I. K.}}}, \bibinfo {author} {\bibnamefont {{White, M.}}}, \bibinfo {author} {\bibnamefont {{White, S. D. M.}}}, \bibinfo {author} {\bibnamefont {{Zacchei, A.}}},\ and\ \bibinfo {author} {\bibnamefont {{Zonca, A.}}},\ }\href {https://doi.org/10.1051/0004-6361/201833910} {\bibfield  {journal} {\bibinfo  {journal} {A\&A}\ }\textbf {\bibinfo {volume} {641}},\ \bibinfo {pages} {A6} (\bibinfo {year} {2020})}\BibitemShut {NoStop}%
\bibitem [{\citenamefont {Mangano}\ \emph {et~al.}(2011{\natexlab{a}})\citenamefont {Mangano}, \citenamefont {Miele}, \citenamefont {Pastor}, \citenamefont {Pisanti},\ and\ \citenamefont {Sarikas}}]{mangano2011constraining}%
  \BibitemOpen
  \bibfield  {author} {\bibinfo {author} {\bibfnamefont {G.}~\bibnamefont {Mangano}}, \bibinfo {author} {\bibfnamefont {G.}~\bibnamefont {Miele}}, \bibinfo {author} {\bibfnamefont {S.}~\bibnamefont {Pastor}}, \bibinfo {author} {\bibfnamefont {O.}~\bibnamefont {Pisanti}},\ and\ \bibinfo {author} {\bibfnamefont {S.}~\bibnamefont {Sarikas}},\ }\href@noop {} {\bibfield  {journal} {\bibinfo  {journal} {Journal of Cosmology and Astroparticle Physics}\ }\textbf {\bibinfo {volume} {2011}}\bibinfo  {number} { (03)},\ \bibinfo {pages} {035}}\BibitemShut {NoStop}%
\bibitem [{\citenamefont {Navas}\ \emph {et~al.}(2024)\citenamefont {Navas} \emph {et~al.}}]{ParticleDataGroup:2024cfk}%
  \BibitemOpen
\bibfield  {number} {  }\bibfield  {author} {\bibinfo {author} {\bibfnamefont {S.}~\bibnamefont {Navas}} \emph {et~al.} (\bibinfo {collaboration} {Particle Data Group}),\ }\href {https://doi.org/10.1103/PhysRevD.110.030001} {\bibfield  {journal} {\bibinfo  {journal} {Phys. Rev. D}\ }\textbf {\bibinfo {volume} {110}},\ \bibinfo {pages} {030001} (\bibinfo {year} {2024})}\BibitemShut {NoStop}%
\bibitem [{\citenamefont {Johns}\ \emph {et~al.}(1996)\citenamefont {Johns}, \citenamefont {Ellis},\ and\ \citenamefont {Lattimer}}]{Johns:1996ht}%
  \BibitemOpen
  \bibfield  {author} {\bibinfo {author} {\bibfnamefont {S.~M.}\ \bibnamefont {Johns}}, \bibinfo {author} {\bibfnamefont {P.~J.}\ \bibnamefont {Ellis}},\ and\ \bibinfo {author} {\bibfnamefont {J.~M.}\ \bibnamefont {Lattimer}},\ }\href {https://doi.org/10.1086/178212} {\bibfield  {journal} {\bibinfo  {journal} {Astrophys. J.}\ }\textbf {\bibinfo {volume} {473}},\ \bibinfo {pages} {1020} (\bibinfo {year} {1996})},\ \Eprint {https://arxiv.org/abs/nucl-th/9604004} {arXiv:nucl-th/9604004} \BibitemShut {NoStop}%
\bibitem [{\citenamefont {de~Forcrand}(2010)}]{deforcrand2010simulatingqcdfinitedensity}%
  \BibitemOpen
  \bibfield  {author} {\bibinfo {author} {\bibfnamefont {P.}~\bibnamefont {de~Forcrand}},\ }\href {https://arxiv.org/abs/1005.0539} {\bibinfo {title} {Simulating qcd at finite density}} (\bibinfo {year} {2010}),\ \Eprint {https://arxiv.org/abs/1005.0539} {arXiv:1005.0539 [hep-lat]} \BibitemShut {NoStop}%
\bibitem [{\citenamefont {Noronha-Hostler}\ \emph {et~al.}(2019)\citenamefont {Noronha-Hostler}, \citenamefont {Parotto}, \citenamefont {Ratti},\ and\ \citenamefont {Stafford}}]{PhysRevC.100.064910}%
  \BibitemOpen
  \bibfield  {author} {\bibinfo {author} {\bibfnamefont {J.}~\bibnamefont {Noronha-Hostler}}, \bibinfo {author} {\bibfnamefont {P.}~\bibnamefont {Parotto}}, \bibinfo {author} {\bibfnamefont {C.}~\bibnamefont {Ratti}},\ and\ \bibinfo {author} {\bibfnamefont {J.~M.}\ \bibnamefont {Stafford}},\ }\href {https://doi.org/10.1103/PhysRevC.100.064910} {\bibfield  {journal} {\bibinfo  {journal} {Phys. Rev. C}\ }\textbf {\bibinfo {volume} {100}},\ \bibinfo {pages} {064910} (\bibinfo {year} {2019})}\BibitemShut {NoStop}%
\bibitem [{\citenamefont {Abuali}\ \emph {et~al.}(2025)\citenamefont {Abuali}, \citenamefont {Bors\'anyi}, \citenamefont {Fodor}, \citenamefont {Jahan}, \citenamefont {Kahangirwe}, \citenamefont {Parotto}, \citenamefont {P\'asztor}, \citenamefont {Ratti}, \citenamefont {Shah},\ and\ \citenamefont {Trabulsi}}]{Abuali:2025tbd}%
  \BibitemOpen
  \bibfield  {author} {\bibinfo {author} {\bibfnamefont {A.}~\bibnamefont {Abuali}}, \bibinfo {author} {\bibfnamefont {S.}~\bibnamefont {Bors\'anyi}}, \bibinfo {author} {\bibfnamefont {Z.}~\bibnamefont {Fodor}}, \bibinfo {author} {\bibfnamefont {J.}~\bibnamefont {Jahan}}, \bibinfo {author} {\bibfnamefont {M.}~\bibnamefont {Kahangirwe}}, \bibinfo {author} {\bibfnamefont {P.}~\bibnamefont {Parotto}}, \bibinfo {author} {\bibfnamefont {A.}~\bibnamefont {P\'asztor}}, \bibinfo {author} {\bibfnamefont {C.}~\bibnamefont {Ratti}}, \bibinfo {author} {\bibfnamefont {H.}~\bibnamefont {Shah}},\ and\ \bibinfo {author} {\bibfnamefont {S.~A.}\ \bibnamefont {Trabulsi}},\ }\href@noop {} {\bibinfo {title} {{A new 4D lattice QCD equation of state: extended density coverage from a generalized $T^\prime$-expansion}}} (\bibinfo {year} {2025}),\ \Eprint {https://arxiv.org/abs/2504.01881} {arXiv:2504.01881 [hep-lat]} \BibitemShut {NoStop}%
\bibitem [{\citenamefont {Jahan}\ and\ \citenamefont {Parotto}(2025)}]{jahan_2025_15123623}%
  \BibitemOpen
  \bibfield  {author} {\bibinfo {author} {\bibfnamefont {J.}~\bibnamefont {Jahan}}\ and\ \bibinfo {author} {\bibfnamefont {P.}~\bibnamefont {Parotto}},\ }\href {https://doi.org/10.5281/zenodo.15123623} {10.5281/zenodo.15123623} (\bibinfo {year} {2025})\BibitemShut {NoStop}%
\bibitem [{\citenamefont {Bazavov}\ \emph {et~al.}(2024)\citenamefont {Bazavov}, \citenamefont {Bollweg}, \citenamefont {Kaczmarek}, \citenamefont {Karsch}, \citenamefont {Mukherjee}, \citenamefont {Petreczky}, \citenamefont {Schmidt},\ and\ \citenamefont {Sharma}}]{Bazavov_2024}%
  \BibitemOpen
  \bibfield  {author} {\bibinfo {author} {\bibfnamefont {A.}~\bibnamefont {Bazavov}}, \bibinfo {author} {\bibfnamefont {D.}~\bibnamefont {Bollweg}}, \bibinfo {author} {\bibfnamefont {O.}~\bibnamefont {Kaczmarek}}, \bibinfo {author} {\bibfnamefont {F.}~\bibnamefont {Karsch}}, \bibinfo {author} {\bibfnamefont {S.}~\bibnamefont {Mukherjee}}, \bibinfo {author} {\bibfnamefont {P.}~\bibnamefont {Petreczky}}, \bibinfo {author} {\bibfnamefont {C.}~\bibnamefont {Schmidt}},\ and\ \bibinfo {author} {\bibfnamefont {S.}~\bibnamefont {Sharma}},\ }\href {https://doi.org/10.1016/j.physletb.2024.138520} {\bibfield  {journal} {\bibinfo  {journal} {Physics Letters B}\ }\textbf {\bibinfo {volume} {850}},\ \bibinfo {pages} {138520} (\bibinfo {year} {2024})}\BibitemShut {NoStop}%
\bibitem [{\citenamefont {Vovchenko}\ \emph {et~al.}(2017)\citenamefont {Vovchenko}, \citenamefont {Gorenstein},\ and\ \citenamefont {Stoecker}}]{PhysRevLett.118.182301}%
  \BibitemOpen
  \bibfield  {author} {\bibinfo {author} {\bibfnamefont {V.}~\bibnamefont {Vovchenko}}, \bibinfo {author} {\bibfnamefont {M.~I.}\ \bibnamefont {Gorenstein}},\ and\ \bibinfo {author} {\bibfnamefont {H.}~\bibnamefont {Stoecker}},\ }\href {https://doi.org/10.1103/PhysRevLett.118.182301} {\bibfield  {journal} {\bibinfo  {journal} {Phys. Rev. Lett.}\ }\textbf {\bibinfo {volume} {118}},\ \bibinfo {pages} {182301} (\bibinfo {year} {2017})}\BibitemShut {NoStop}%
\bibitem [{\citenamefont {Hagedorn}(1965)}]{Hagedorn:1965st}%
  \BibitemOpen
  \bibfield  {author} {\bibinfo {author} {\bibfnamefont {R.}~\bibnamefont {Hagedorn}},\ }\href@noop {} {\bibfield  {journal} {\bibinfo  {journal} {Nuovo Cim. Suppl.}\ }\textbf {\bibinfo {volume} {3}},\ \bibinfo {pages} {147} (\bibinfo {year} {1965})}\BibitemShut {NoStop}%
\bibitem [{\citenamefont {Vovchenko}\ and\ \citenamefont {Stoecker}(2019{\natexlab{a}})}]{Vovchenko:2019pjl}%
  \BibitemOpen
  \bibfield  {author} {\bibinfo {author} {\bibfnamefont {V.}~\bibnamefont {Vovchenko}}\ and\ \bibinfo {author} {\bibfnamefont {H.}~\bibnamefont {Stoecker}},\ }\href {https://doi.org/10.1016/j.cpc.2019.06.024} {\bibfield  {journal} {\bibinfo  {journal} {Comput. Phys. Commun.}\ }\textbf {\bibinfo {volume} {244}},\ \bibinfo {pages} {295} (\bibinfo {year} {2019}{\natexlab{a}})},\ \Eprint {https://arxiv.org/abs/1901.05249} {arXiv:1901.05249 [nucl-th]} \BibitemShut {NoStop}%
\bibitem [{\citenamefont {Vovchenko}\ and\ \citenamefont {Stoecker}(2019{\natexlab{b}})}]{VOVCHENKO2019295}%
  \BibitemOpen
  \bibfield  {author} {\bibinfo {author} {\bibfnamefont {V.}~\bibnamefont {Vovchenko}}\ and\ \bibinfo {author} {\bibfnamefont {H.}~\bibnamefont {Stoecker}},\ }\href {https://doi.org/https://doi.org/10.1016/j.cpc.2019.06.024} {\bibfield  {journal} {\bibinfo  {journal} {Computer Physics Communications}\ }\textbf {\bibinfo {volume} {244}},\ \bibinfo {pages} {295} (\bibinfo {year} {2019}{\natexlab{b}})}\BibitemShut {NoStop}%
\bibitem [{\citenamefont {Castorina}\ \emph {et~al.}(2012)\citenamefont {Castorina}, \citenamefont {França}, \citenamefont {Lattanzi}, \citenamefont {Lesgourgues}, \citenamefont {Mangano}, \citenamefont {Melchiorri},\ and\ \citenamefont {Pastor}}]{Castorina_2012}%
  \BibitemOpen
  \bibfield  {author} {\bibinfo {author} {\bibfnamefont {E.}~\bibnamefont {Castorina}}, \bibinfo {author} {\bibfnamefont {U.}~\bibnamefont {França}}, \bibinfo {author} {\bibfnamefont {M.}~\bibnamefont {Lattanzi}}, \bibinfo {author} {\bibfnamefont {J.}~\bibnamefont {Lesgourgues}}, \bibinfo {author} {\bibfnamefont {G.}~\bibnamefont {Mangano}}, \bibinfo {author} {\bibfnamefont {A.}~\bibnamefont {Melchiorri}},\ and\ \bibinfo {author} {\bibfnamefont {S.}~\bibnamefont {Pastor}},\ }\bibfield  {journal} {\bibinfo  {journal} {Physical Review D}\ }\textbf {\bibinfo {volume} {86}},\ \href {https://doi.org/10.1103/physrevd.86.023517} {10.1103/physrevd.86.023517} (\bibinfo {year} {2012})\BibitemShut {NoStop}%
\bibitem [{\citenamefont {Mangano}\ \emph {et~al.}(2011{\natexlab{b}})\citenamefont {Mangano}, \citenamefont {Miele}, \citenamefont {Pastor}, \citenamefont {Pisanti},\ and\ \citenamefont {Sarikas}}]{Mangano_2011}%
  \BibitemOpen
  \bibfield  {author} {\bibinfo {author} {\bibfnamefont {G.}~\bibnamefont {Mangano}}, \bibinfo {author} {\bibfnamefont {G.}~\bibnamefont {Miele}}, \bibinfo {author} {\bibfnamefont {S.}~\bibnamefont {Pastor}}, \bibinfo {author} {\bibfnamefont {O.}~\bibnamefont {Pisanti}},\ and\ \bibinfo {author} {\bibfnamefont {S.}~\bibnamefont {Sarikas}},\ }\href {https://doi.org/10.1088/1475-7516/2011/03/035} {\bibfield  {journal} {\bibinfo  {journal} {Journal of Cosmology and Astroparticle Physics}\ }\textbf {\bibinfo {volume} {2011}}\bibinfo  {number} { (03)},\ \bibinfo {pages} {035–035}}\BibitemShut {NoStop}%
\bibitem [{\citenamefont {Pastor}(2010)}]{SergioPastor_2010}%
  \BibitemOpen
\bibfield  {number} {  }\bibfield  {author} {\bibinfo {author} {\bibfnamefont {S.}~\bibnamefont {Pastor}},\ }\href {https://doi.org/10.1088/1742-6596/203/1/012053} {\bibfield  {journal} {\bibinfo  {journal} {Journal of Physics: Conference Series}\ }\textbf {\bibinfo {volume} {203}},\ \bibinfo {pages} {012053} (\bibinfo {year} {2010})}\BibitemShut {NoStop}%
\bibitem [{\citenamefont {Mangano}\ \emph {et~al.}(2012{\natexlab{b}})\citenamefont {Mangano}, \citenamefont {Miele}, \citenamefont {Pastor}, \citenamefont {Pisanti},\ and\ \citenamefont {Sarikas}}]{Mangano_2012}%
  \BibitemOpen
  \bibfield  {author} {\bibinfo {author} {\bibfnamefont {G.}~\bibnamefont {Mangano}}, \bibinfo {author} {\bibfnamefont {G.}~\bibnamefont {Miele}}, \bibinfo {author} {\bibfnamefont {S.}~\bibnamefont {Pastor}}, \bibinfo {author} {\bibfnamefont {O.}~\bibnamefont {Pisanti}},\ and\ \bibinfo {author} {\bibfnamefont {S.}~\bibnamefont {Sarikas}},\ }\href {https://doi.org/10.1016/j.physletb.2012.01.015} {\bibfield  {journal} {\bibinfo  {journal} {Physics Letters B}\ }\textbf {\bibinfo {volume} {708}},\ \bibinfo {pages} {1–5} (\bibinfo {year} {2012}{\natexlab{b}})}\BibitemShut {NoStop}%
\bibitem [{\citenamefont {Barenboim}\ \emph {et~al.}(2017)\citenamefont {Barenboim}, \citenamefont {Kinney},\ and\ \citenamefont {Park}}]{PhysRevD.95.043506}%
  \BibitemOpen
  \bibfield  {author} {\bibinfo {author} {\bibfnamefont {G.}~\bibnamefont {Barenboim}}, \bibinfo {author} {\bibfnamefont {W.~H.}\ \bibnamefont {Kinney}},\ and\ \bibinfo {author} {\bibfnamefont {W.-I.}\ \bibnamefont {Park}},\ }\href {https://doi.org/10.1103/PhysRevD.95.043506} {\bibfield  {journal} {\bibinfo  {journal} {Phys. Rev. D}\ }\textbf {\bibinfo {volume} {95}},\ \bibinfo {pages} {043506} (\bibinfo {year} {2017})}\BibitemShut {NoStop}%
\bibitem [{\citenamefont {Oldengott}\ and\ \citenamefont {Schwarz}(2017{\natexlab{b}})}]{Oldengott_2017}%
  \BibitemOpen
  \bibfield  {author} {\bibinfo {author} {\bibfnamefont {I.~M.}\ \bibnamefont {Oldengott}}\ and\ \bibinfo {author} {\bibfnamefont {D.~J.}\ \bibnamefont {Schwarz}},\ }\href {https://doi.org/10.1209/0295-5075/119/29001} {\bibfield  {journal} {\bibinfo  {journal} {EPL (Europhysics Letters)}\ }\textbf {\bibinfo {volume} {119}},\ \bibinfo {pages} {29001} (\bibinfo {year} {2017}{\natexlab{b}})}\BibitemShut {NoStop}%
\bibitem [{\citenamefont {Wygas}\ \emph {et~al.}(2018{\natexlab{b}})\citenamefont {Wygas}, \citenamefont {Oldengott}, \citenamefont {B\"odeker},\ and\ \citenamefont {Schwarz}}]{PhysRevLett.121.201302}%
  \BibitemOpen
  \bibfield  {author} {\bibinfo {author} {\bibfnamefont {M.~M.}\ \bibnamefont {Wygas}}, \bibinfo {author} {\bibfnamefont {I.~M.}\ \bibnamefont {Oldengott}}, \bibinfo {author} {\bibfnamefont {D.}~\bibnamefont {B\"odeker}},\ and\ \bibinfo {author} {\bibfnamefont {D.~J.}\ \bibnamefont {Schwarz}},\ }\href {https://doi.org/10.1103/PhysRevLett.121.201302} {\bibfield  {journal} {\bibinfo  {journal} {Phys. Rev. Lett.}\ }\textbf {\bibinfo {volume} {121}},\ \bibinfo {pages} {201302} (\bibinfo {year} {2018}{\natexlab{b}})}\BibitemShut {NoStop}%
\bibitem [{\citenamefont {Middeldorf-Wygas}\ \emph {et~al.}(2022)\citenamefont {Middeldorf-Wygas}, \citenamefont {Oldengott}, \citenamefont {B\"odeker},\ and\ \citenamefont {Schwarz}}]{PhysRevD.105.123533}%
  \BibitemOpen
  \bibfield  {author} {\bibinfo {author} {\bibfnamefont {M.~M.}\ \bibnamefont {Middeldorf-Wygas}}, \bibinfo {author} {\bibfnamefont {I.~M.}\ \bibnamefont {Oldengott}}, \bibinfo {author} {\bibfnamefont {D.}~\bibnamefont {B\"odeker}},\ and\ \bibinfo {author} {\bibfnamefont {D.~J.}\ \bibnamefont {Schwarz}},\ }\href {https://doi.org/10.1103/PhysRevD.105.123533} {\bibfield  {journal} {\bibinfo  {journal} {Phys. Rev. D}\ }\textbf {\bibinfo {volume} {105}},\ \bibinfo {pages} {123533} (\bibinfo {year} {2022})}\BibitemShut {NoStop}%
\end{thebibliography}%

\end{document}